\documentclass[epj]{svjour}
\usepackage{graphics}
\begin{document}
\title{Radiation Hardness of MALTA2 Monolithic CMOS Sensors on Czochralski Substrates}

\author{Milou van Rijnbach\inst{1,2} \and Dumitru Vlad Berlea\inst{3} \and Valerio Dao\inst{1} \and Martin Ga\v{z}i\inst{4} \and Phil Allport\inst{5} \and Ignacio Asensi Tortajada\inst{1} \and Prafulla Behera\inst{6} \and Daniela Bortoletto\inst{4} \and Craig Buttar\inst{7} \and Florian Dachs\inst{1} \and Ganapati Dash\inst{6} \and Dominik Dobrijevi\'c\inst{1,8} \and Lucian Fasselt\inst{3} \and Leyre Flores Sanz de Acedo\inst{1} \and Andrea Gabrielli\inst{1} \and Vicente Gonz\'alez\inst{9} \and Giuliano Gustavino\inst{1} \and Pranati Jana\inst{6} \and Heinz Pernegger\inst{1} \and Petra Riedler\inst{1} \and Heidi Sandaker\inst{2} \and Carlos Solans S\'anchez\inst{1} \and Walter Snoeys\inst{1} \and Tomislav Suligoj\inst{8} \and Marcos V\'azquez N\'u\~nez\inst{1,9} \and Anusree Vijay\inst{6} \and Julian Weick\inst{1,10} \and Steven Worm\inst{3} \and Abdelhak M. Zoubir\inst{10}.
}
\mail{milou.van.rijnbach@cern.ch}
\institute{CERN, Geneva, Switzerland \and University of Oslo, Oslo, Norway  \and DESY, Zeuthen, Germany \and University of Oxford, Oxford, United Kingdom \and University of Birmingham, Birmingham, United Kingdom \and Indian Institute of Technology Madras, Chennai, India  \and University of Glasgow, Glasgow, United Kingdom \and University of Zagreb, Zagreb, Croatia \and Universitat de València, València, Spain \and Technische Universität Darmstadt, Darmstadt, Germany}
\date{Received: date / Revised version: date}
%
\abstract{
MALTA2 is the latest full-scale prototype of the MALTA family of Depleted Monolithic Active Pixel Sensors (DMAPS) produced in Tower Semiconductor 180 nm CMOS technology. In order to comply with the requirements of High Energy Physics (HEP) experiments, various process modifications and front-end changes have been implemented to achieve low power consumption, reduce Random Telegraph Signal (RTS) noise, and optimise the charge collection geometry. Compared to its predecessors, MALTA2 targets the use of a high-resistivity, thick Czochralski (Cz) substrates in order to demonstrate radiation hardness in terms of detection efficiency and timing resolution up to 3$\times$10$^{15}$ 1 MeV $\mathrm{n_{eq}/{cm}^2}$ with backside metallisation to achieve good propagation of the bias voltage. This manuscript shows the results that were obtained with non-irradiated and irradiated MALTA2 samples on Cz substrates from the CERN SPS test beam campaign from 2021-2023 using the MALTA telescope.
\PACS{
      {PACS-key}{discribing text of that key}   \and
      {PACS-key}{discribing text of that key}
     } 
} 
\maketitle
\section{History of the MALTA Family}

\noindent Monolithic CMOS pixel sensors provide several advantages over hybrid pixel sensors for High Energy Physics (HEP) experiments. These include a high level of integration, scalability, and low power consumption. Additionally, monolithic sensor designs featuring a small collection electrode can be operated with low noise due to the reduced capacitance. These sensors are expected to play a crucial role in future particle physics experiments and have been the focus of various R\&D projects over the past decade \cite{colaleo20212021}. They are now approaching a mature stage of development as they are being implemented in several HEP experiments, such as ALICE \cite{mager2016alpide}. \\

\noindent The MALTA chip was developed for potential use in the ATLAS experiment at the High Luminosity LHC (HL-LHC) upgrade and possible integration in other future high-energy physics experiments. Its design targets radiation hardness for fluences $>$10$^{15}$ 1 MeV $\mathrm{n_{eq}/{cm}^2}$  (NIEL) and 100 Mrad (TID), thin CMOS sensors with high granularity, high hit-rate capability ($>$100 MHz/cm$^2$), and fast response time (40 MHz). The MALTA matrix is made up of 512$\times$512 pixels, each with a pixel pitch of 36.4\textmu m and a small octagonal-shaped collection electrode (2 \textmu m diameter) that results in low noise and low power dissipation (10 mW/cm$^2$ digital at 100 MHz/cm$^2$ and 70 mW/cm$^2$ analog power). The asynchronous readout sends hit information directly from the pixel to the periphery through 37 parallel output signals with a 2 ns output signal length. This avoids distributing high-frequency clock signals across the matrix, minimising analog-digital cross-talk and power consumption. Pixels are organised in 2$\times$8 groups and hits are sent to a common reference pulse generator within each group. The reference pulse is added to the pixel and group address (16-bit and 5-bit respectively) and hits are distributed through two parallel 22-bit wide buses, one for even groups and the other for odd groups. This distinction reduces cross-talk on the hit address bus as adjacent groups cannot share the same bus \cite{pernegger2023malta}.\\

\begin{figure*}
\centering
\resizebox{0.5\textwidth}{!}{
\includegraphics{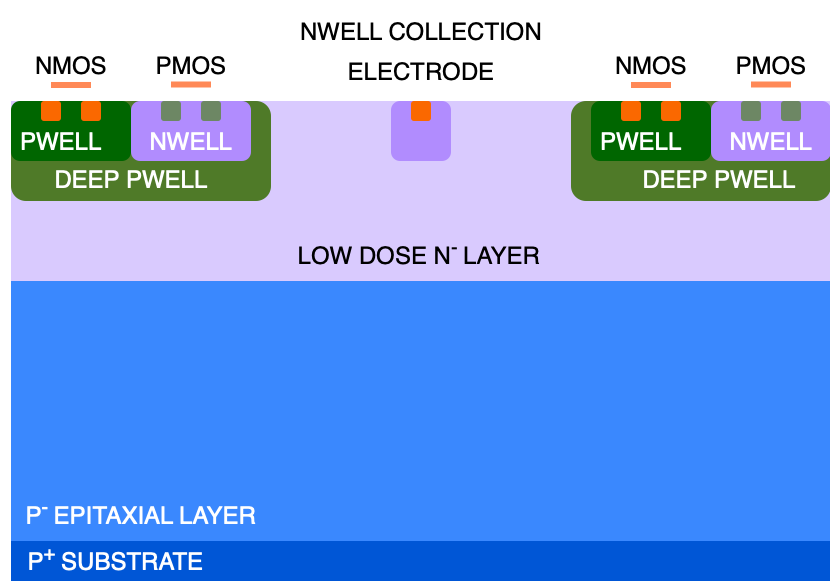}
}
\resizebox{1\textwidth}{!}{
\includegraphics{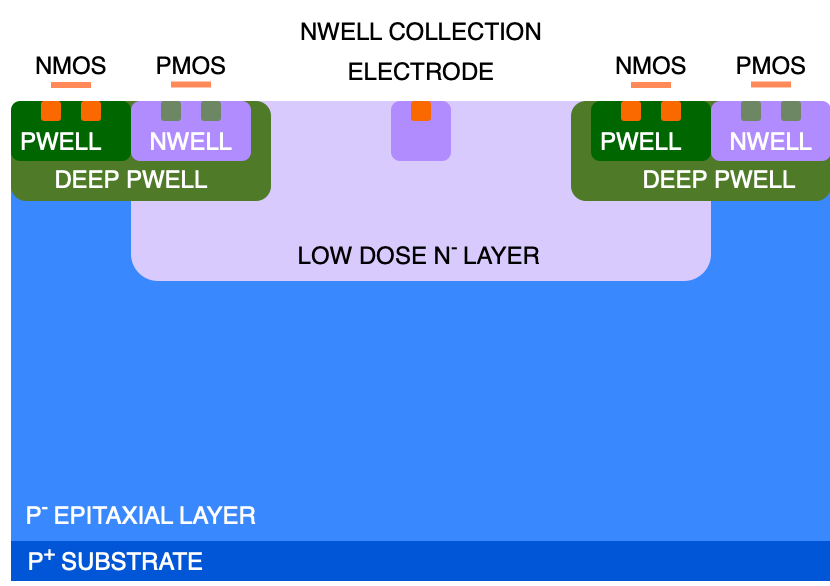}
\includegraphics{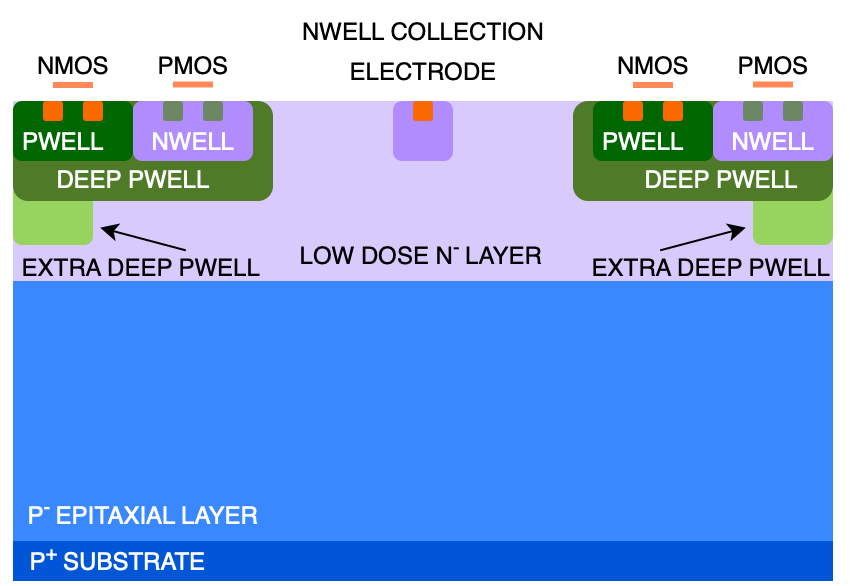}
}
\caption{Cross sections of the process modifications of the Tower Semiconductor 180 nm CMOS imaging technology. Top image shows the standard modified process (STD) where an n$^-$ layer is introduced on top of the p-type substrate. Bottom left image shows the process modification where a gap in the low dose n$^-$ layer is introduced (NGAP). The bottom right image shows the process modification with an extra deep p-well located under the deep p-well (XDPW). Images are not drawn to scale and are adapted from Ref.\cite{munker2019simulations}.}
\label{fig:crosssection}
\end{figure*}

\noindent The MALTA sensor is fabricated in a modified Tower Semiconductor 180 nm CMOS imaging technology using three different pixel flavours, illustrated in Figure \ref{fig:crosssection}. The standard modified process (STD) introduces a low dose n$^-$ layer across the full pixel matrix on top of the p-type substrate, the second process modification includes a design with a gap in the n$^-$ layer (NGAP), and the third pixel design features an additional deep p-well implant (XDPW) \cite{munker2019simulations}. The NGAP and XDPW pixel flavours have demonstrated to be particularly appealing for radiation hard detectors, as they have shown improved detection efficiency in the pixel corners \cite{dyndal2020mini}. For MALTA, these pixel flavours are produced on high-resistivity p-type epitaxial substrates and high-resistivity p-type Czochralski (Cz) substrates. The results presented in Ref.\cite{pernegger2023malta} demonstrate the effectiveness of MALTA on high-resistivity Cz substrates as an alternative wafer material. Sensors with Cz substrates have shown the capability to achieve a larger depletion volume, leading to a significantly amplified ionization charge signal in comparison to MALTA sensors on epitaxial substrates.\\

\noindent This paper presents the results that were achieved with the next generation MALTA chip, MALTA2, on Cz wafers. The paper will discuss the design modifications that were implemented with respect to its predecessor and their implications on the performance. This will be followed by a discussion on the importance of a good backside contact for the Cz wafers. Thereafter, the test beam set-up for sample characterisation will be discussed, including a summary of the various MALTA2 samples used in this study. Finally, test beam results at the Super Proton Synchrotron (SPS) at CERN will be presented on the radiation hardness of MALTA2 Cz samples.

\section{MALTA2}

 MALTA2 is the second generation detector of the MALTA family. The main objectives of the new variant are expanding the radiation hardness of the design towards higher Displacement Damage Doses (DDD) \cite{dyndal2020mini}, achieving a uniform in-pixel charge collection and lowering the Random Telegraph Signal (RTS) noise in the sensor front-end \cite{piro20221}. MALTA2 is approximately half the size of the MALTA sensor, with a matrix of $224\times512$ pixels ($9\times18$ mm$^2$). It inherits the asynchronous readout of the previous generation, but implements modifications in both the slow control and front-end. These modifications underwent validation through a small-scale demonstrator, called the mini-MALTA \cite{dyndal2020mini}.
 
\subsection{Slow Control and Front-End Changes}

The original MALTA detector's slow control system utilised an Ethernet-like protocol. However, in the mini-MALTA detector and subsequently in MALTA2, a more efficient design was realised by incorporating a shift register \cite{de2022latest}. This shift register-based slow control not only enabled more efficient configuration of the chip but also demonstrated successful operation in other silicon-based DMAPS technologies \cite{caicedo2022development}, thereby facilitating a more reliable implementation. \\

\noindent Several changes in the pixel front-end have been implemented in order to achieve the goals of the MALTA2 variant, discussed in Ref.\cite{piro20221}. In order to capitalise on the additional charge generated in thicker substrates, an open-loop amplification was implemented. This leads to a compact frontend design with lower noise and higher speed of the circuit suitable for a small electrode pixel layout.  Multiple cascode transistors were implemented to both enhance the overall gain of the front-end and to decouple key transistors from the input and output lines. This allows the resizing of the transistor gates without heavy penalties on the total capacitance. Large gate areas have been implemented for the input and amplification stage transistor in order to limit the RTS \cite{piro20221}. The improved front-end brings improvements in terms of lower noise and the elimination of RTS, shown in Ref.\cite{van2022radiation}. \\

\begin{figure*}
\centering
\resizebox{0.85\textwidth}{!}{
\hspace*{-1.2cm} 
\includegraphics{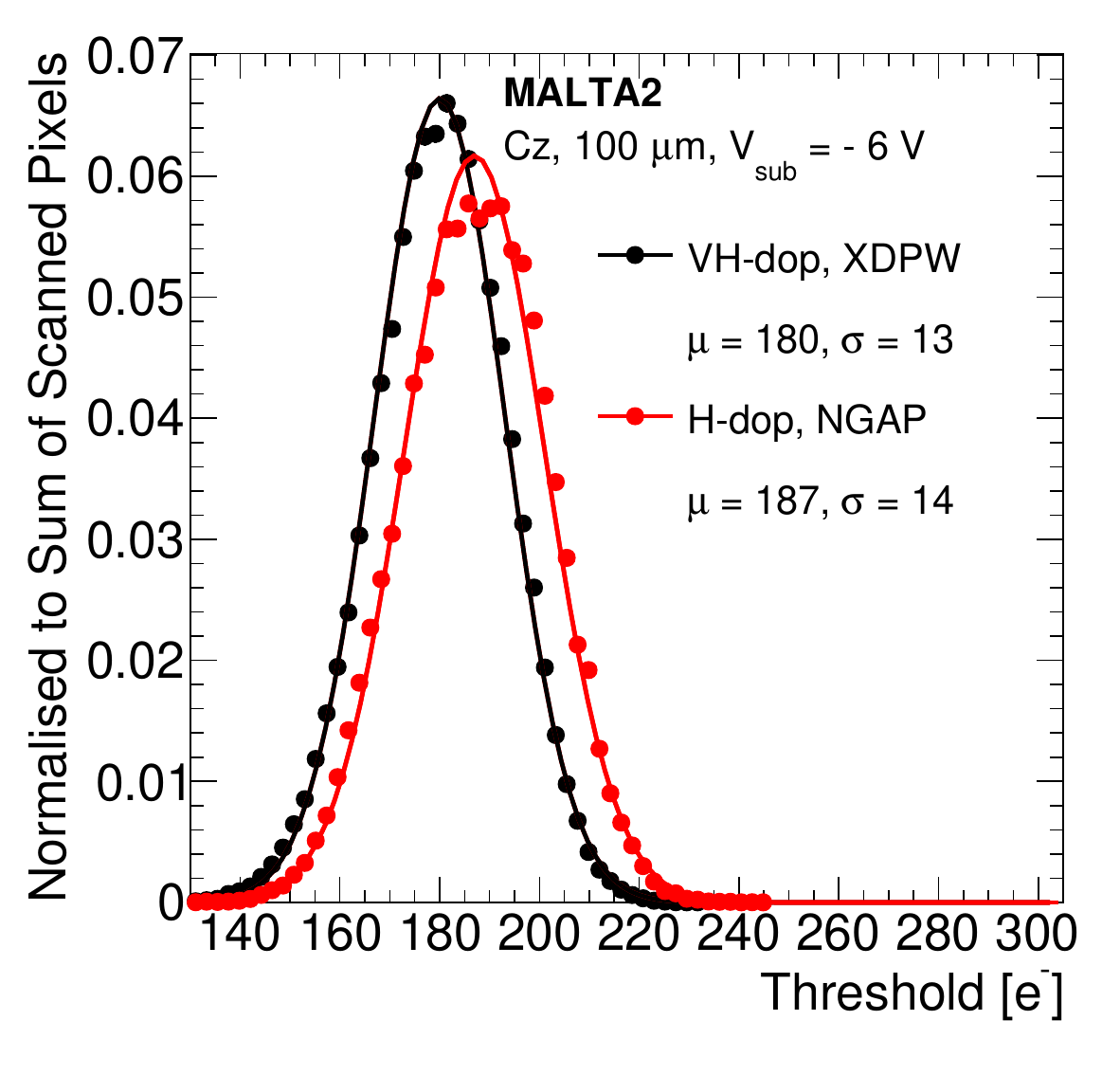}
\hspace*{4cm} 
\includegraphics{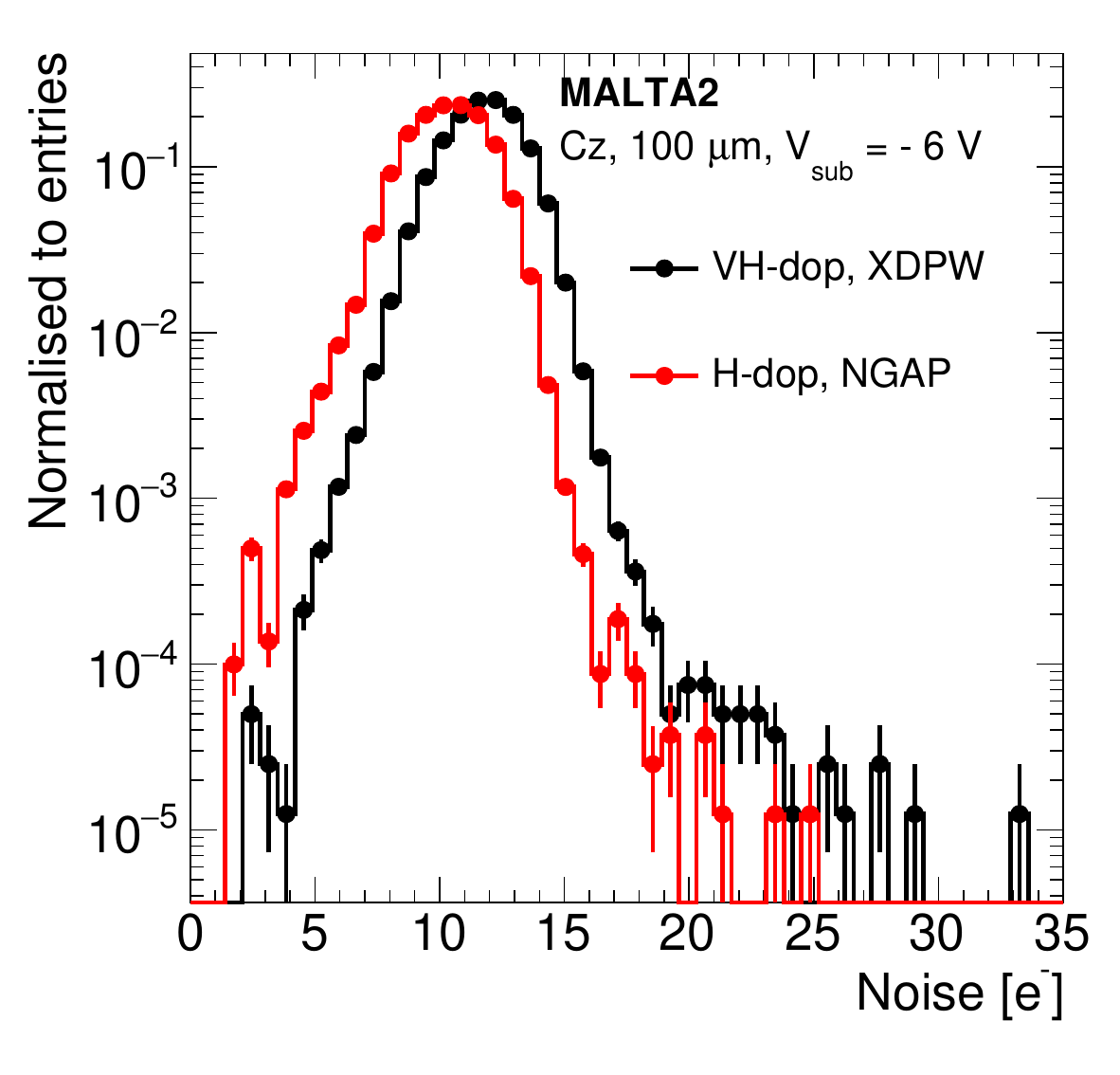}
}
\resizebox{1\textwidth}{!}{
\includegraphics{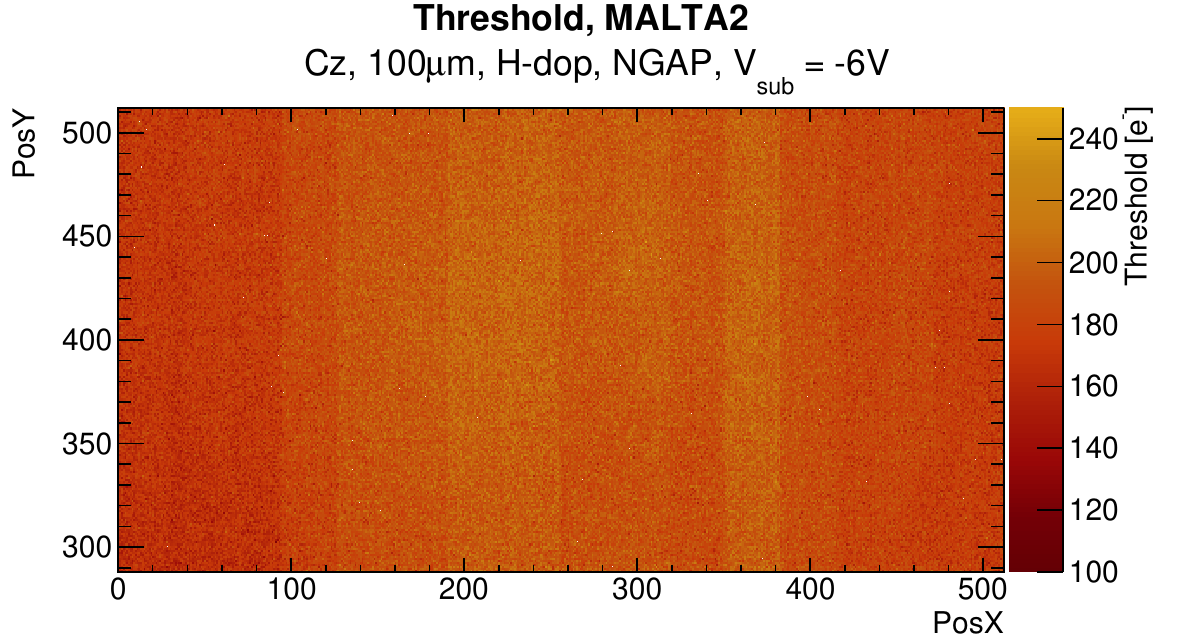}
\includegraphics{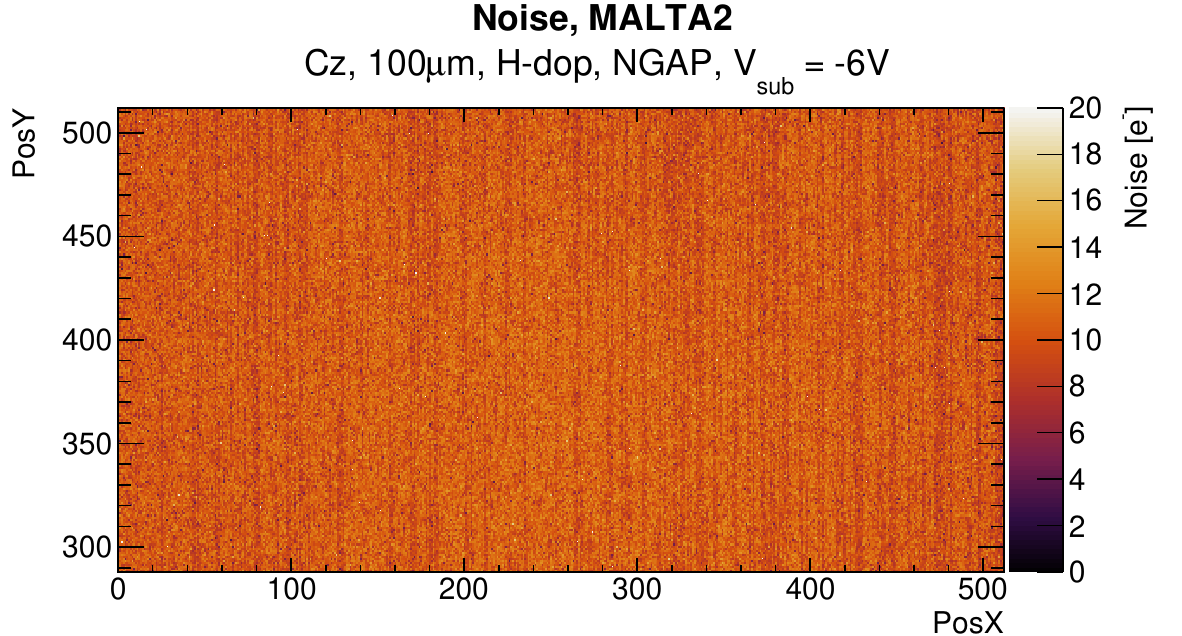}
}
\resizebox{1\textwidth}{!}{
\includegraphics{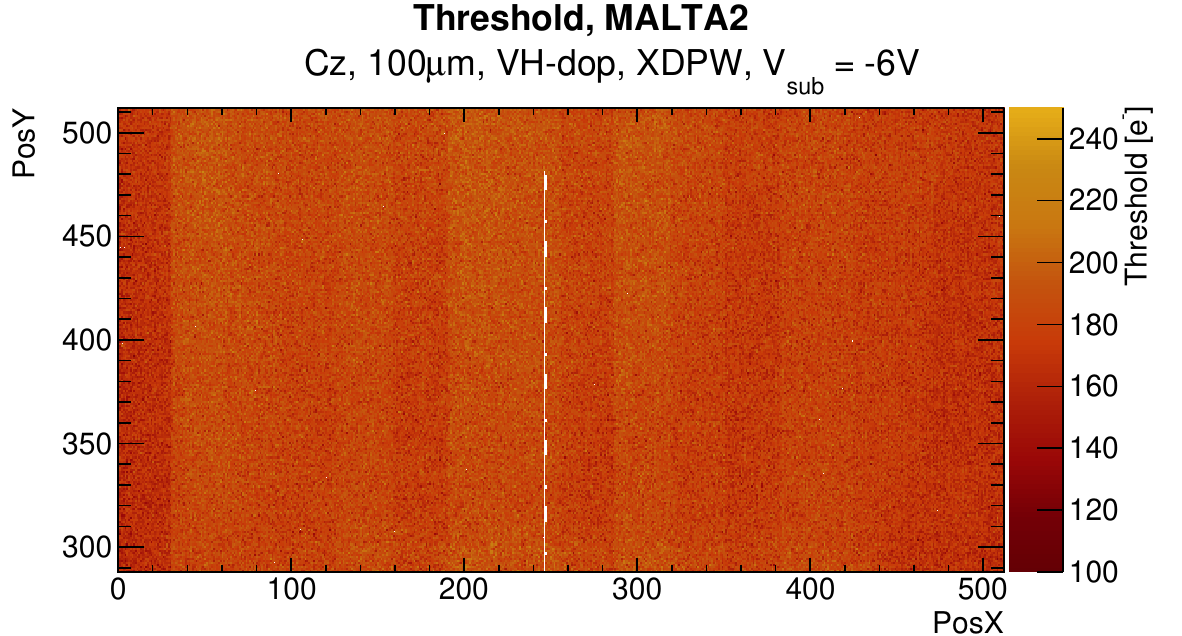}
\includegraphics{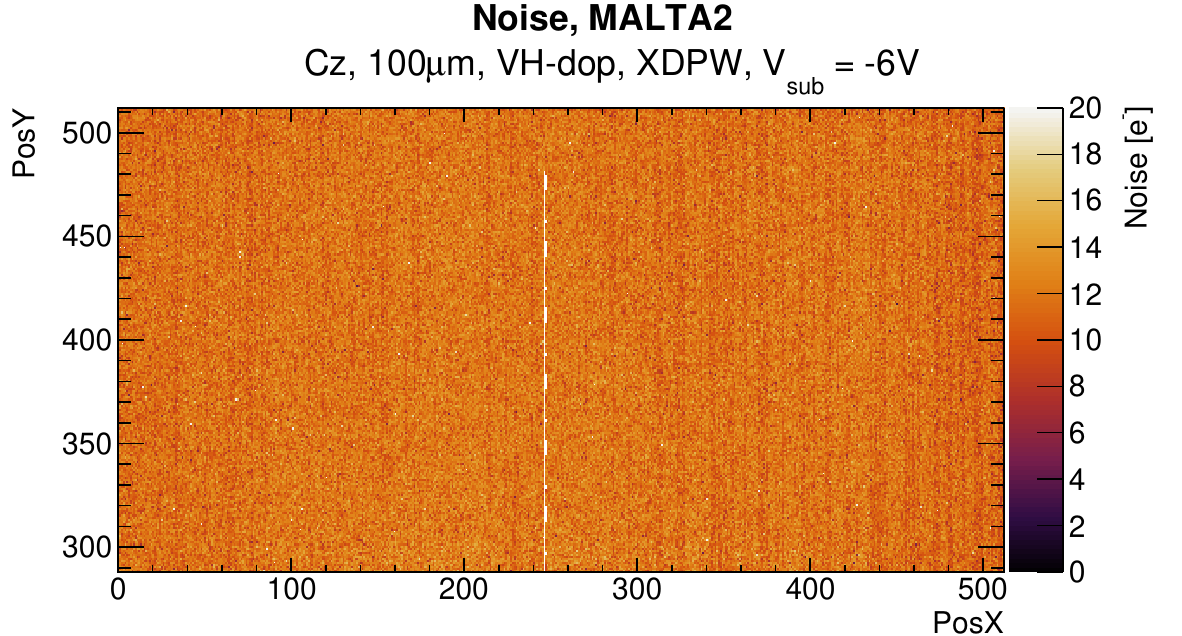}
}
\caption{Threshold distribution (top left image) and noise distribution (top right image), of non-irradiated  MALTA2 (Cz, 100 \textmu m) NGAP, high doping of n$^-$ layer (in red) and XDPW, very high doping (in black) at -6 V SUB bias. Threshold corresponds to $\sim$180 e$^-$. The cut-off for the noise distribution at 2 e$^-$ is correlated to the granularity of the noise scan. Additionally the corresponding 2D distribution for the entire matrix of the high doping sample are shown (middle images) and for the very high doping sample (bottom images).  }
\label{fig:Thr_noise}
\end{figure*}

\noindent Figure \ref{fig:Thr_noise} highlights the threshold and noise distributions for non-irradiated MALTA2 samples with high and very high doping of the n$^-$ layer. For the same threshold (within the expected $10\%$ dispersion), a larger noise is observed for the MALTA2 with the very high doping of the n$^-$ layer. This effect is correlated to the increase in capacitance due to the thinner depletion zone around the collection electrode for an higher doped n$^-$ layer \cite{Dort:2813457}. Additionally, a low threshold value can be applied across the whole sensor. The threshold is uniformly distributed in the column (Y) direction, whereas a small variation in threshold along the row (X) direction of the matrix is observed, which is correlated to the front-end biasing scheme. Power pads are distributed on the left and right side of the matrix, leading to an incremental horizontal power voltage drop \cite{piro20221}. Additionally, the noise is distributed uniformly across the entire matrix. A very small noise tail indicates that there is a minor contribution from RTS to the total noise. Additionally, it has been shown that both TID and DDD have a relatively small impact on the threshold and noise of the MALTA2 sensor \cite{piro20221}. Manual tuning of the threshold has been shown to counteract the changes in average efficiency due to radiation damage.

\subsection{Process Modifications for Radiation Hard Monolithic CMOS Sensors}
\label{sec:promod}

As highlighted in Ref.\cite{dyndal2020mini}, on the testing of the mini-MALTA demonstrator, the addition of a lightly-doped (compared to the doping concentration of the collection electrode) n$^-$ layer across the whole area of the pixel is efficient at extending the depletion region in the lateral direction. It allows for the electron-hole pairs that are generated in the active depth, inside the sensing volume, to move through drift. This feature holds higher importance for sensors that are expected to be irradiated up to large DDD.  In the process of displacement damage, an incident particle or photon can dislodge a silicon atom from its lattice site and hereby create deep level acceptor and donor traps. Trapping sites have a larger impact on charges that travel only through diffusion to the collection electrode \cite{Dort:2813457}.\\

\noindent For the MALTA2 sensors, the doping concentration of the deep n$^-$ layer has been modified to gauge its impact on the performance and radiation hardness of the sensor. The effect of this doping concentration has been studied in TCAD simulations \cite{Dort:2813457}. For higher doping concentrations, the sensor capacitance is affected due to a decrease of the depletion thickness around the collection electrode. The increase in the doping of the n$^-$ layer is expected to compensate for the possible type inversion of the n$^-$ layer, especially prevalent at high fluence levels due to the acceptor creation \cite{pernegger2023malta}. \\

\noindent Initially, MAPS became a promising technology for tracking sensors due to the possibility of, among others, adapting existing CMOS imaging technology for HEP applications \cite{SNOEYS201941}. Close partnership with the chip foundry has allowed to use high resistivity (3-4 k$\mathrm{\Omega}$cm) Cz wafers as active substrates. MALTA-Cz with 100 and 300 \textmu m thick substrates have shown promising results \cite{pernegger2023malta} in terms of better radiation hardness and larger cluster size. In general, for the MALTA2 variant, Cz samples have a larger active depth (compared to other silicon growth techniques) that need a large substrate voltage to achieve full depletion. Only irradiated samples can be biased to large voltages ($>50$~V) due to the increased punch through voltage that was identified through IV studies in Ref.\cite{pernegger2023malta}. Cz substrate needs additional considerations in order to achieve a good backside propagation of the voltage, especially for irradiated samples.

\subsection{Backside Processing}

\begin{figure}
\centering
\resizebox{0.5\textwidth}{!}{
\includegraphics{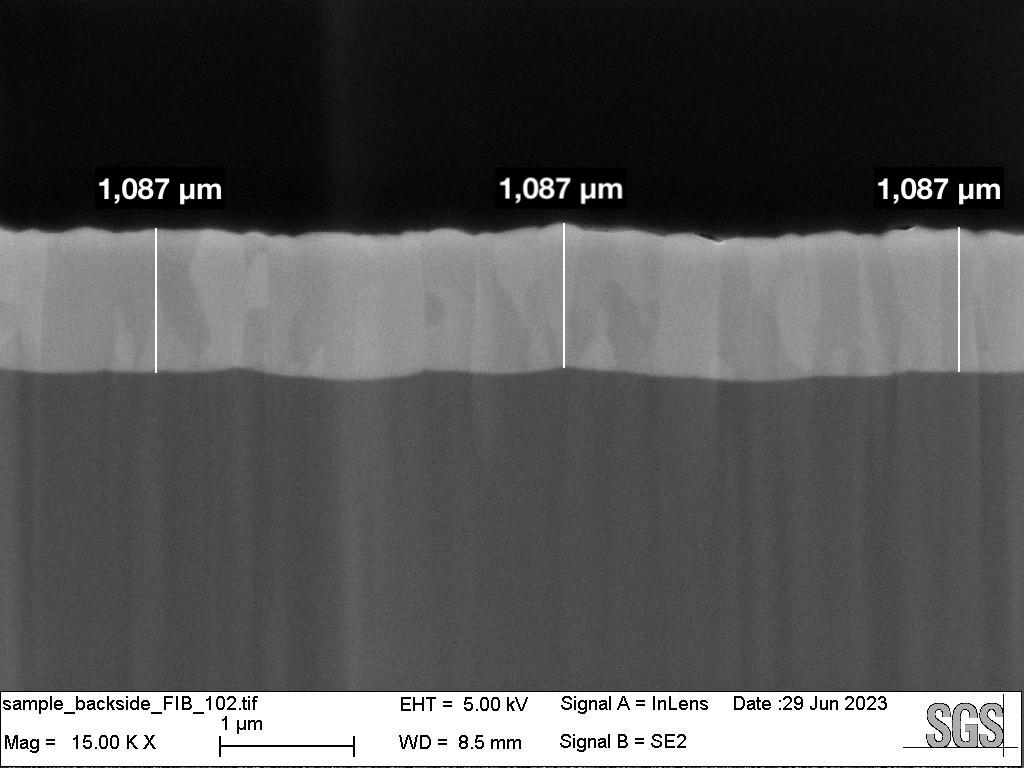}
}
\caption{Cross-sectional SEM image of a MALTA2 sample with backside metallisation. The light grey area indicates the 1 \textmu m thick Aluminium layer.}
\label{fig:IBS}
\end{figure}

\noindent By default (for non-irradiated samples), the voltage propagation from the Printed Circuit Board (PCB) to the chip proceeds through the left and right side of the matrix through an electrically conductive film, Staystik\textsuperscript \textregistered \cite{staystik}, which is placed along the borders of the hole that exists in the PCB. For irradiated Cz samples, backside metallisation was explored to achieve a good propagation of the substrate voltage, as the default method was found not to propagate the bias voltage uniformly across the whole chip at elevated bias voltages. The backside metallisation process is performed by Ion Beam Services (IBS) and consists of four consecutive operations: thinning, implantation, annealing, and aluminium deposition. First, the Cz wafer is thinned down using a TAIKO thinning process \cite{disco} with a membrane of 100~\textmu m through plasma etching. This step is performed in order to achieve the desired thickness of the substrate while maintaining mechanical stability of the wafers during the process. For the p-type implantation step, boron is used as the dopant. After implantation, the samples are placed in a Rapid Thermal Anneal (RTA) chamber on a silicon support which facilitates a pyrometer in order to measure the temperature. In the last step, a 1 \textmu m +/- 10\% thick aluminium layer is deposited on the backside of the sample. Figure~\ref{fig:IBS} presents a Scanning Electron Microscope (SEM) image of a cross-section of a MALTA2 sample with backside metallisation.

\section{Test Beam Set-up for Sample Characterisation}

\noindent Between 2021 and 2023, a dedicated test beam campaign was conducted at the Super Proton Synchrotron (SPS) facility of CERN with a 180 GeV hadron beam, using the MALTA telescope to characterise the samples. The main objective of this campaign was to carry out the characterisation of MALTA2 Cz samples with respect to their radiation tolerance and timing performance, although other variables were also evaluated.

\subsection{Test Beam with MALTA Telescope}

\noindent The MALTA telescope, a custom pixel telescope, comprises of six MALTA tracking planes (consisting of epitaxial samples and two Cz samples) and a scintillator for timing reference located behind the telescope planes. It facilitates the testing of two Devices Under Test (DUTs) simultaneously and it features a custom cold-box to host irradiated samples in a dry environment at cold temperatures. The trigger system of the telescope is fully configurable, enabling triggering on coincidence between the telescope planes and the scintillator. More information regarding the architecture and performance of the MALTA telescope is available in Ref.\cite{van2023performance}. For the alignment, track reconstruction, and offline analysis of the test beam data, the software package Proteus is used \cite{kiehn}. The observables that are used to characterise the DUT, i.e. cluster size, hit detection efficiency, and timing resolution, are defined in the sections where the respective results are discussed. \\

\noindent Measurements for non-irradiated samples are performed at room temperature, i.e. 20$^\circ$C, and measurements with irradiated samples at -20$^\circ$C. All the irradiated samples examined in this study underwent the backside metallisation (back-metal) process. The quoted threshold values are extracted by using theoretical values for the injection capacitance.

\subsection{Sample Collection}

The MALTA2 sensor has been fabricated with various process modifications. In order to facilitate the understanding of the results presented in Chapter \ref{Sec:Eff} and \ref{Sec:Timing}, the various design and operational parameters are explained in more detail below. Given the limited number of available samples and the extensive range of operating parameters, only a subset of the parameters have been studied isolatedly. An overview of the MALTA2 samples studied in this work is provided in Table \ref{tabel1}. \\

\begin{table*}[!ht]
\centering
    \begin{tabular}{ |  p{3.5cm} | p{1.2cm} | p{1.2cm}  | p{1.6cm} | p{1.6cm} | p{1.6cm} | p{1.6cm} | p{1.6cm} |}
           \hline
            \multicolumn{8}{c}{Overview of MALTA2 Samples} \\ \hline
Fluence & 0 & 0 & 0 & 1$\times$10$^{15}$ & 2$\times$10$^{15}$ & 3$\times$10$^{15}$ & 3$\times$10$^{15}$ \\ 
Sensor flavour & NGAP & XDPW & XDPW & XDPW & XDPW & XDPW & XDPW \\ 
Total thickness  [\textmu m] & 300 & 100 & 100 & 100  & 100 & 100  & 100 \\ 
Doping level of n$^-$ layer & H & VH  & H & H & H  & H & VH \\ 
Backside post-processing & None & None & back-metal & back-metal & back-metal & back-metal & back-metal   \\ \hline
          \end{tabular}
 \caption{Overview of the main specifications of the MALTA2 samples studied in this work. For every sample the fluence level, sensor flavour, and thickness have been indicated. The doping level of the n$^-$ layer is indicated as H (high) and VH (very high). All irradiated samples and one non-irradiated sample underwent backside metallisation (back-metal) during post-processing.}
 \label{tabel1}
 \end{table*}

\noindent\textbf{Process Modification}\\

\noindent For the MALTA2 Cz samples in this study, the available flavours are NGAP and XDPW. As shown in Ref.\cite{pernegger2023malta}, the performance after irradiation between these two flavours is comparable. The results presented in the following sections will not discuss the difference in performance between these flavours.\\

\noindent\textbf{Doping Level of Continuous n$^-$ layer}\\

\noindent Two distinct doping concentrations are present for the continuous n$^-$ layer, denoted as high (H-dop) and very high (VH-dop) doping. The doping concentration of the deep n$^-$ layer plays a crucial role in achieving high radiation tolerance. It should be pointed out that the naming convention (high and very high) refers to their relative difference in implantation dose, approximately 70\%, and does not refer to the absolute doping concentration itself. \\

\noindent\textbf{Sensor Thickness}\\

\noindent All MALTA2 Cz wafers are thinned down to 300 \textmu m or 100 \textmu m to minimise the material budget. For all samples, the quoted thickness includes a metal stack on top of the sensor, which is approximately 10 \textmu m thick. \\

\noindent\textbf{Fluence Levels}\\

\noindent Results will be presented before and after neutron irradiation. Sensors have been irradiated with neutrons to 1, 2, and 3$\times$10$^{15}$ 1 MeV $\mathrm{n_{eq}/{cm}^2}$  at the Triga reactor at the Institute Jožef Stefan, Ljubljana, Slovenia \cite{ambrovzivc2017computational}\cite{snoj2012computational}. The irradiated samples have been subjected to an annealing process at room temperature for several days. However, no samples were subject to long-term reverse annealing at elevated temperatures.  \\

\noindent\textbf{Operating Bias Voltage}\\

\noindent The voltage of the p-well is fixed at -6 V. In order to generate a larger drift field in samples with a thick substrate, the reverse bias  ($\mathrm{V_{sub}}$) is increased. The operational limit of the bias voltage is restricted by the compliance level set to 2 mA in order to protect the electronics. This in turn implies that as the samples have different design variables, i.e. fluence or thickness, the maximum operational substrate bias voltage is sample dependent.

\section{Efficiency and Cluster Size}
\label{Sec:Eff}

\noindent Achieving efficiencies close to $100\%$ is crucial for tracking detectors, particularly in the context of HEP experiments that utilise multiple sensors in tracking modules \cite{colaleo20212021}. During test beam campaigns, the hit detection efficiency for the MALTA2 samples is calculated as the number of matched clusters on the DUT over the total number of reconstructed tracks. A matched cluster is found by associating hit clusters on the DUT to a track, which should be found within 80 \textmu m. Due to the very large statistics, a small statistical error is recorded for both the hit efficiency and cluster size measurements. During data acquisition, several noisy pixels are masked, still their contribution to the total efficiency is taken into account. 

\subsection{Before Displacement Damage}

\begin{figure*}
\centering
\resizebox{1\textwidth}{!}{
\includegraphics{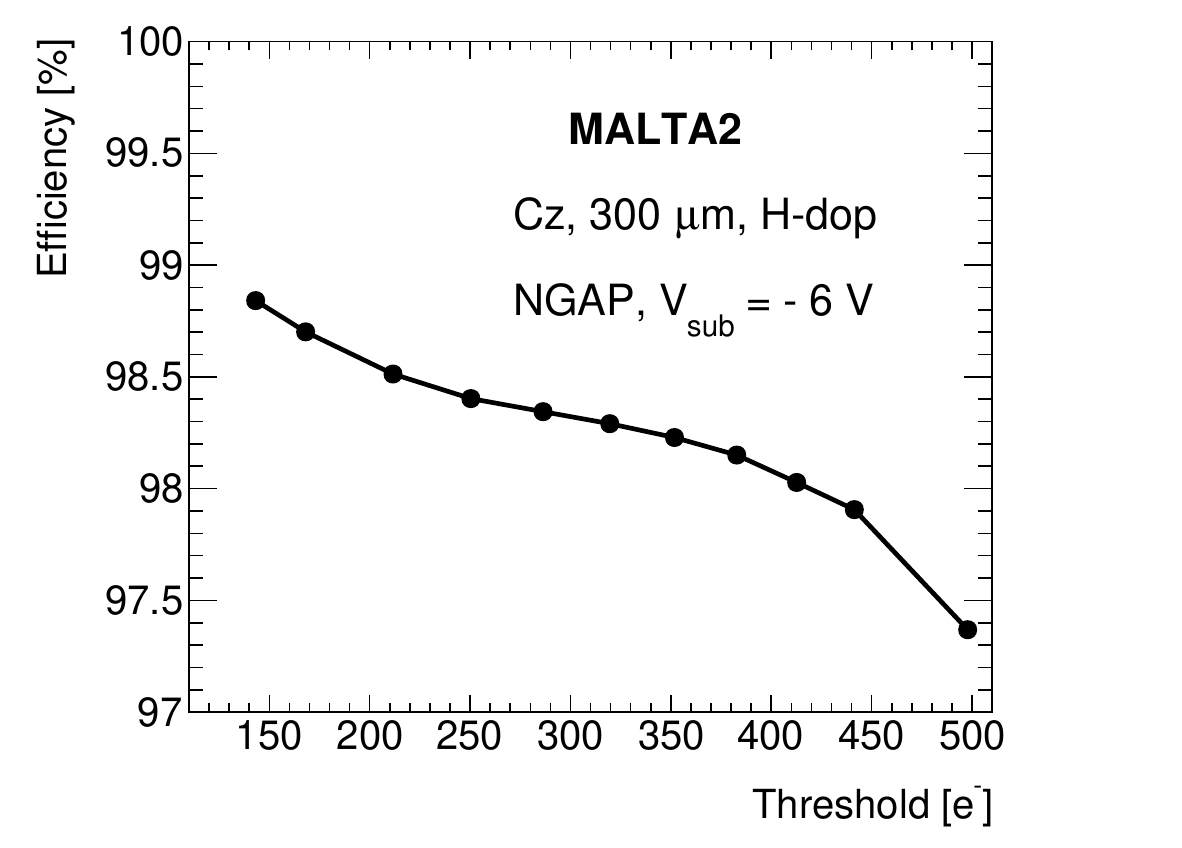}
\includegraphics{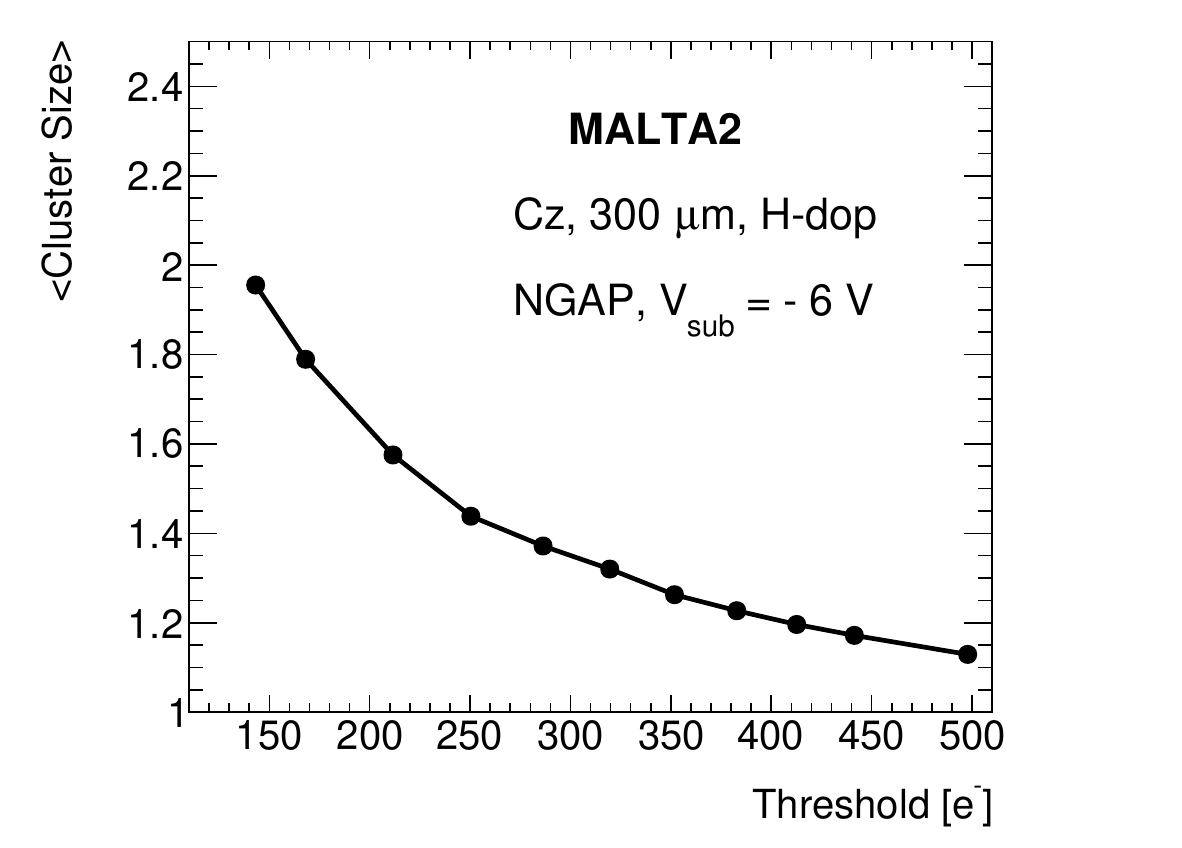}
}
\resizebox{1\textwidth}{!}{
\includegraphics{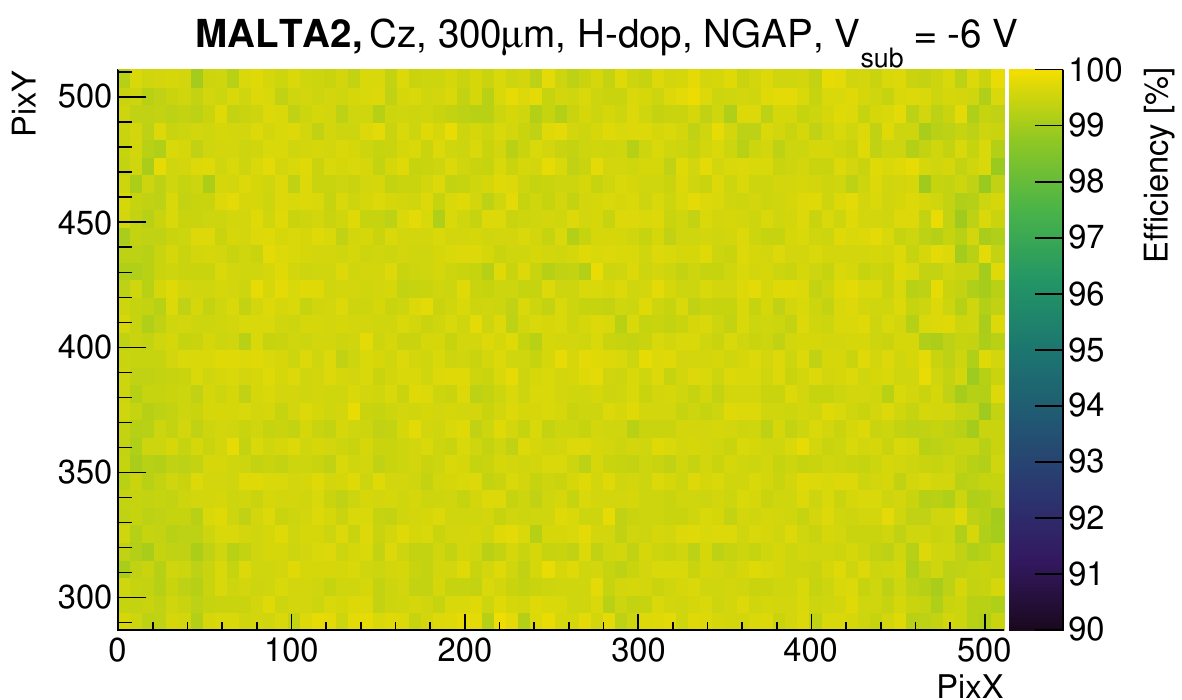}
\includegraphics{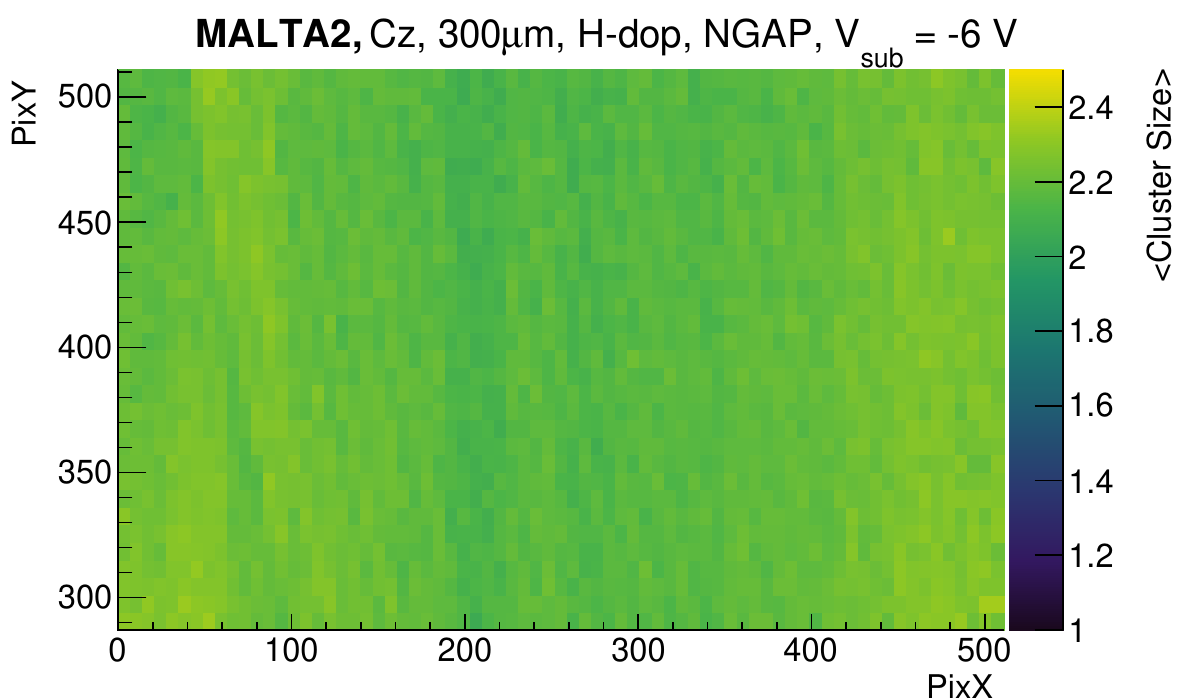}
}
\caption{Average efficiency (top left) and cluster size (top right) as a function of threshold. The bottom images show their respective 2D map at an operating threshold corresponding to 150 e$^-$. Results are shown for a non-irradiated MALTA2 sample (Cz, NGAP, 300 \textmu m thick, high doping of n$^-$ layer) at -6 V. }
\label{fig:Cl_Eff_noirad}
\end{figure*}

\noindent Non-irradiated MALTA2 samples have been characterised in terms of efficiency and cluster size performance for multiple threshold configurations at the default substrate voltage setting of $-6$~V. Figure \ref{fig:Cl_Eff_noirad} highlights the increase in these figures of merit with a decrease in threshold, due to the enhanced detection of hits with low signal amplitude, such as pixel charge sharing and pixel corner hits. With operational threshold settings below $\sim$250 e$^-$, the detection of a larger number of shared charge events yields a large increase in cluster size and a relatively smaller increase in efficiency. Additionally, both a uniform efficiency and cluster size response are observed across the entire matrix. A similar trend in efficiency and cluster size versus threshold was observed between samples with high versus very high doping of the n$^-$ layer.  


\subsection{After Displacement Damage}

\begin{figure*}
\centering
\hspace*{3.5cm} 
\resizebox{0.9\textwidth}{!}{
\includegraphics{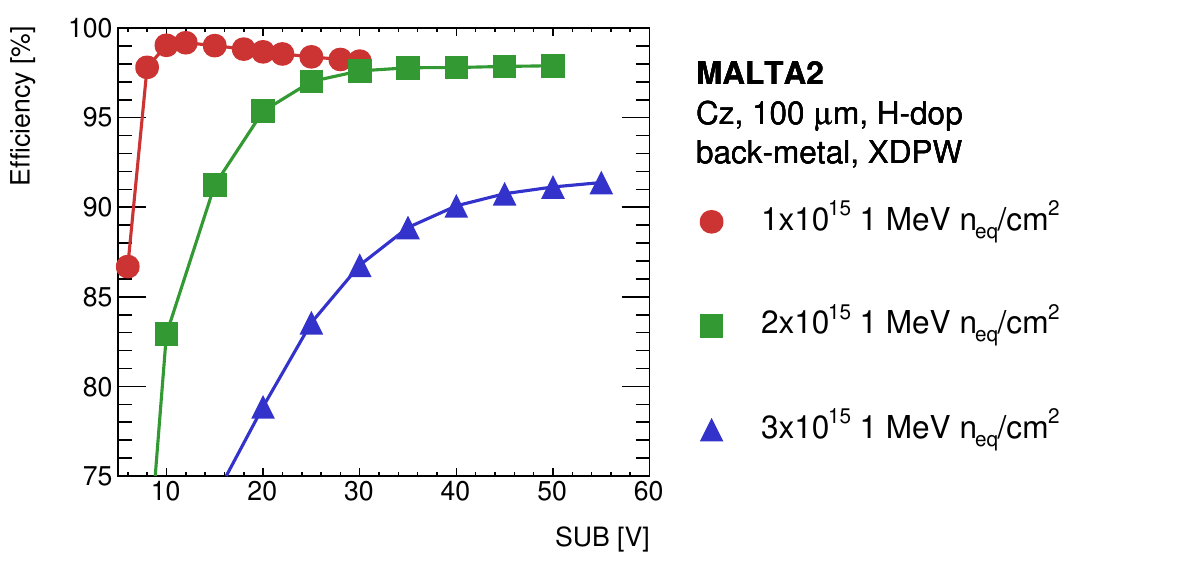}
}
\caption{Average efficiency versus bias voltage for three MALTA2 samples (XDPW, high doping n$^-$ layer, 100 \textmu m thick, and backside metallisation). The samples are irradiated to 1, 2, and 3$\times$10$^{15}$ 1 MeV $\mathrm{n_{eq}/{cm}^2}$  and the results are shown at best performing operating threshold, corresponding to 240, 260, and 120 e$^-$, respectively.}
\label{fig:EffvsSUB}
\end{figure*}

\begin{figure*}
\centering
\resizebox{0.6\textwidth}{!}{
\includegraphics{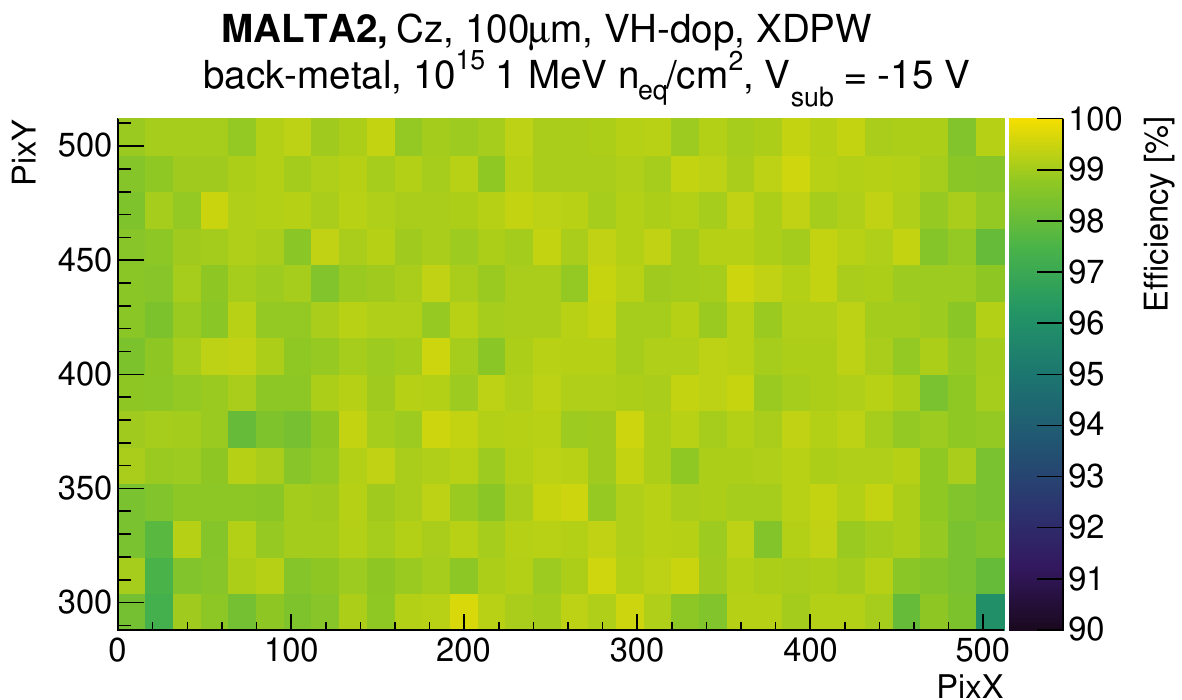}
}
\caption{2D Efficiency map of the entire matrix of a MALTA2 sample (Cz, XDPW, 100 \textmu m thick, high doping of n$^-$ layer and backside metallisation) irradiated to 1$\times$10$^{15}$ 1 MeV $\mathrm{n_{eq}/{cm}^2}$  and operated at $-15$ V. The average efficiency is 99\% and the operating threshold corresponds to 240 e$^-$.}
\label{fig:Eff2D_irad}
\end{figure*}

\begin{figure*}
\centering
\hspace*{3.5cm} 
\resizebox{0.9\textwidth}{!}{
\includegraphics{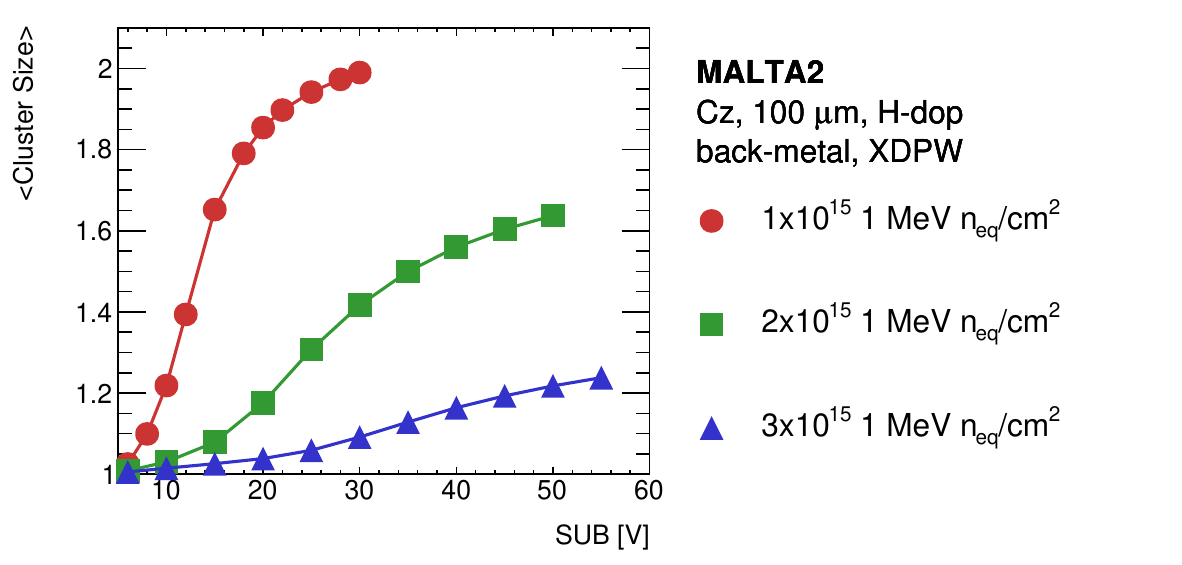}
}
\caption{Average cluster size versus bias voltage for three MALTA2 samples (XDPW, high doping n$^-$ layer, 100 \textmu m thick, and backside metallisation). The samples are irradiated to 1, 2, and 3$\times$10$^{15}$ 1 MeV $\mathrm{n_{eq}/{cm}^2}$  and the results are shown at best performing operating threshold, corresponding to 240, 260, and 120 e$^-$, respectively.}
\label{fig:ClusSizevsSUB}
\end{figure*}

\noindent In order to mitigate the effect of charge trapping, the drift volume can be extended by increasing the substrate voltage. Figure \ref{fig:EffvsSUB} shows the increase in average efficiency with substrate voltage for three MALTA2 samples irradiated at different fluences and operated at similar threshold configurations. All samples have the same doping concentration of the n$^-$ layer, denoted as high doping. For the  1$\times$10$^{15}$ 1 MeV $\mathrm{n_{eq}/{cm}^2}$  sample, a large efficiency is achieved ($99\%$) at relatively low substrate voltage ($<10$ V). With elevated radiation damage (2$\times$10$^{15}$ 1 MeV $\mathrm{n_{eq}/{cm}^2}$ ), a higher substrate voltage is required to obtain high efficiency. This is motivated by the fact that due to the accumulation of radiation damage, the effective doping of the substrate increases, leading to a lower total resistivity, requiring a larger bias voltage to achieve similar depletion depths. For the 3$\times$10$^{15}$ 1 MeV $\mathrm{n_{eq}/{cm}^2}$  sample, only an efficiency of $90\%$ is recovered. Additionally, Figure \ref{fig:Eff2D_irad} shows that for a MALTA2 sample irradiated to 1$\times$10$^{15}$ 1 MeV $\mathrm{n_{eq}/{cm}^2}$ , the efficiency is uniformly distributed over the entire matrix. \\

\noindent Figure \ref{fig:ClusSizevsSUB} shows the increase in average cluster size with substrate voltage for the same irradiated MALTA2 samples presented in Figure \ref{fig:EffvsSUB}. The cluster size increases for higher substrate voltages, due to the enhanced charge sharing effect between pixels for larger active depths in the Cz substrate. With elevated accumulated DDD, a large cluster size is recovered by increasing the substrate voltage. This effect can be explained both by the reduction of the active depth at the same substrate voltage for highly irradiated samples, and by the accumulation of radiation induced charge traps that limit charge sharing in between pixels. The impact of the substrate voltage on the cluster size for the various irradiation doses is similar to the efficiency response, with a large cluster size ($>2$) reconstructed at relatively low substrate voltage (-$20$~V) for the sample irradiated to 1$\times$10$^{15}$ 1 MeV $\mathrm{n_{eq}/{cm}^2}$ . Lower cluster sizes are found at higher substrate voltages for the samples irradiated to 2 and 3$\times$10$^{15}$ 1 MeV $\mathrm{n_{eq}/{cm}^2}$ . This is mainly due to the limited efficiencies of these samples for several substrate bias configurations, which limits the impact of the detected pixel charge sharing.

\subsection{Effect of the Doping Level of the n$^-$ Layer} 

\begin{figure*}
\centering
\hspace*{3.5cm} 
\resizebox{0.9\textwidth}{!}{
\includegraphics{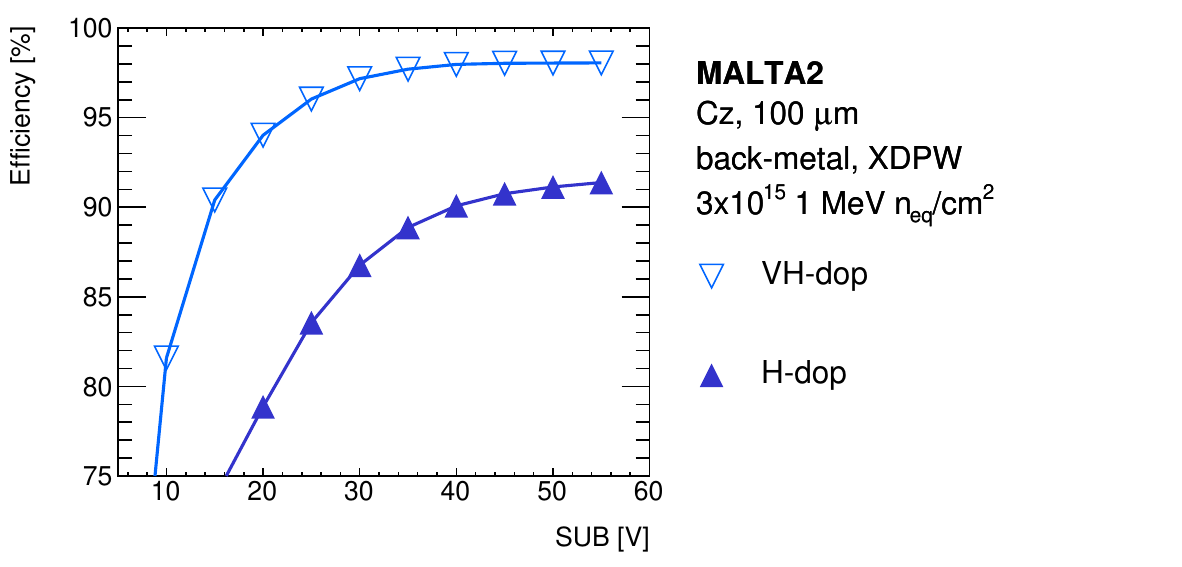}
}
\caption{Average efficiency versus bias voltage for two MALTA2 samples (Cz, XDPW, 100 \textmu m thick, and backside metallisation) irradiated to 3$\times$10$^{15}$ 1 MeV $\mathrm{n_{eq}/{cm}^2}$ . The samples differ in the doping level of the n$^-$ layer, i.e. high and very high, and the results are shown for best performing operational threshold, corresponding to 120 and 110 e$^-$, respectively.}
\label{fig:highvsveryhigh}
\end{figure*}

\begin{figure*}
\centering
\resizebox{0.9\textwidth}{!}{
\includegraphics{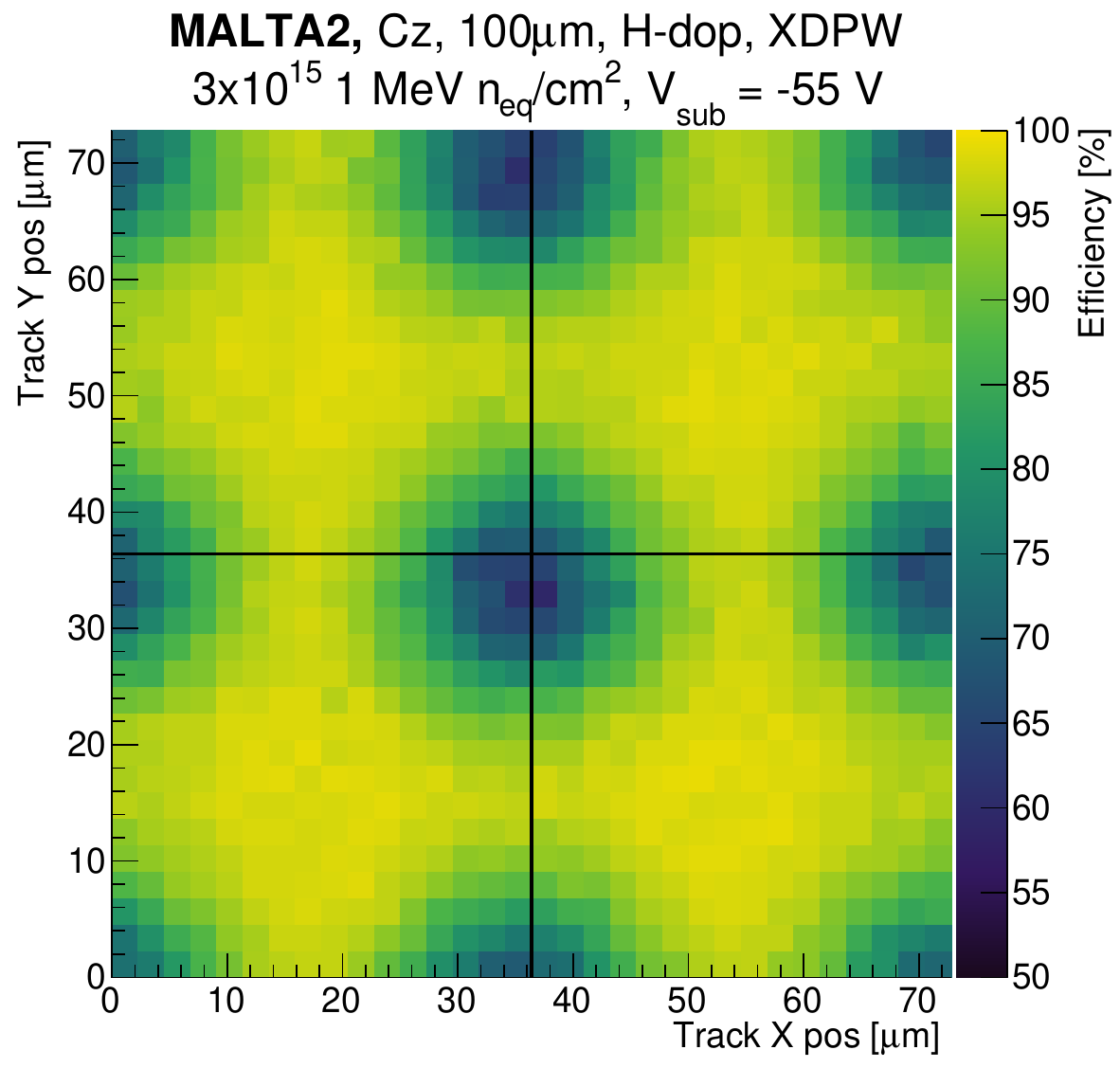}
\includegraphics{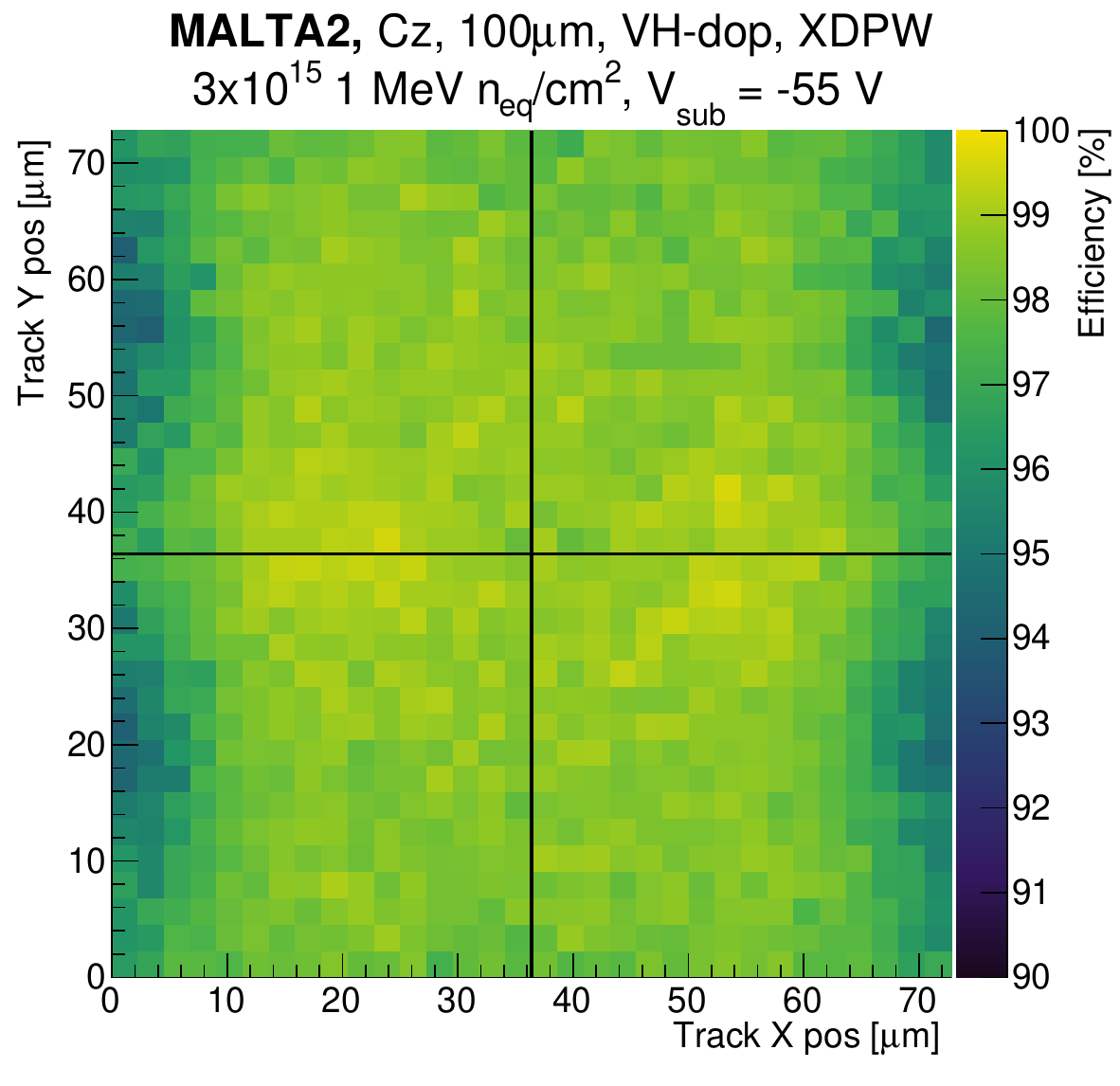}
}
\caption{In-pixel efficiency projected over a 2$\times$2 pixel matrix for two MALTA2 samples (Cz, XDPW, 100 \textmu m thick, and backside metallisation) irradiated to 3$\times$10$^{15}$ 1 MeV $\mathrm{n_{eq}/{cm}^2}$  and operated at $-55$ V. The samples differ in the doping level of the n$^-$ layer, i.e. high (left) and very high (right), and the results are shown for best performing operational threshold, corresponding to 120 and 110 e$^-$, respectively. Note the difference in Z-axis scale. }
\label{fig:Eff_Inpix}
\end{figure*}

A possible explanation for the lower efficiency achieved for the sample irradiated to 3$\times$10$^{15}$ 1 MeV $\mathrm{n_{eq}/{cm}^2}$ in Figure \ref{fig:EffvsSUB} is the doping level of the n$^-$ layer. In order to verify this, a MALTA2 sample with very high doping of the n$^-$ layer has been irradiated to the same fluence level. Figure \ref{fig:highvsveryhigh} shows the impact of the doping concentration on the efficiency versus substrate bias voltage for two samples irradiated to 3$\times$10$^{15}$ 1 MeV $\mathrm{n_{eq}/{cm}^2}$ . As can be seen, a much higher efficiency ($>97\%$) is achieved at a lower substrate voltage for the sample with the very high doping of the n$^-$ layer. The reasoning behind this observation stems from the hypothesis that the sample featuring the very high doping of the n$^-$ layer is able to compensate more for the possible type inversion of the n$^-$ layer, due to the larger acceptor creation at elevated fluence levels. Additionally, it implies that complete depletion might not be accomplished for samples subjected to substantial fluences, particularly when the doping level of the n$^-$ layer is not high enough, as the (deep) p-well and p-type substrate are not sufficiently separated by the n$^-$ layer.\\

\noindent Figure \ref{fig:Eff_Inpix} shows the in-pixel efficiency after 3$\times$10$^{15}$ 1 MeV $\mathrm{n_{eq}/{cm}^2}$  irradiation for two samples with different doping concentrations of the n$^-$ layer: high and very high doping. The efficiency loss for the high doping sample originates from the pixel corners, where the electric field configuration becomes in-efficient in collecting the generated charge. An improvement is observed for the sample with very high doping of the n$^-$ layer, due to the preservation of the initial (non-irradiated) pixel electrical field configuration at high irradiation doses.

\subsection{Operational Window After Displacement Damage}
\label{Sec:operwind}

\begin{figure*}
\centering
\resizebox{0.6\textwidth}{!}{
\includegraphics{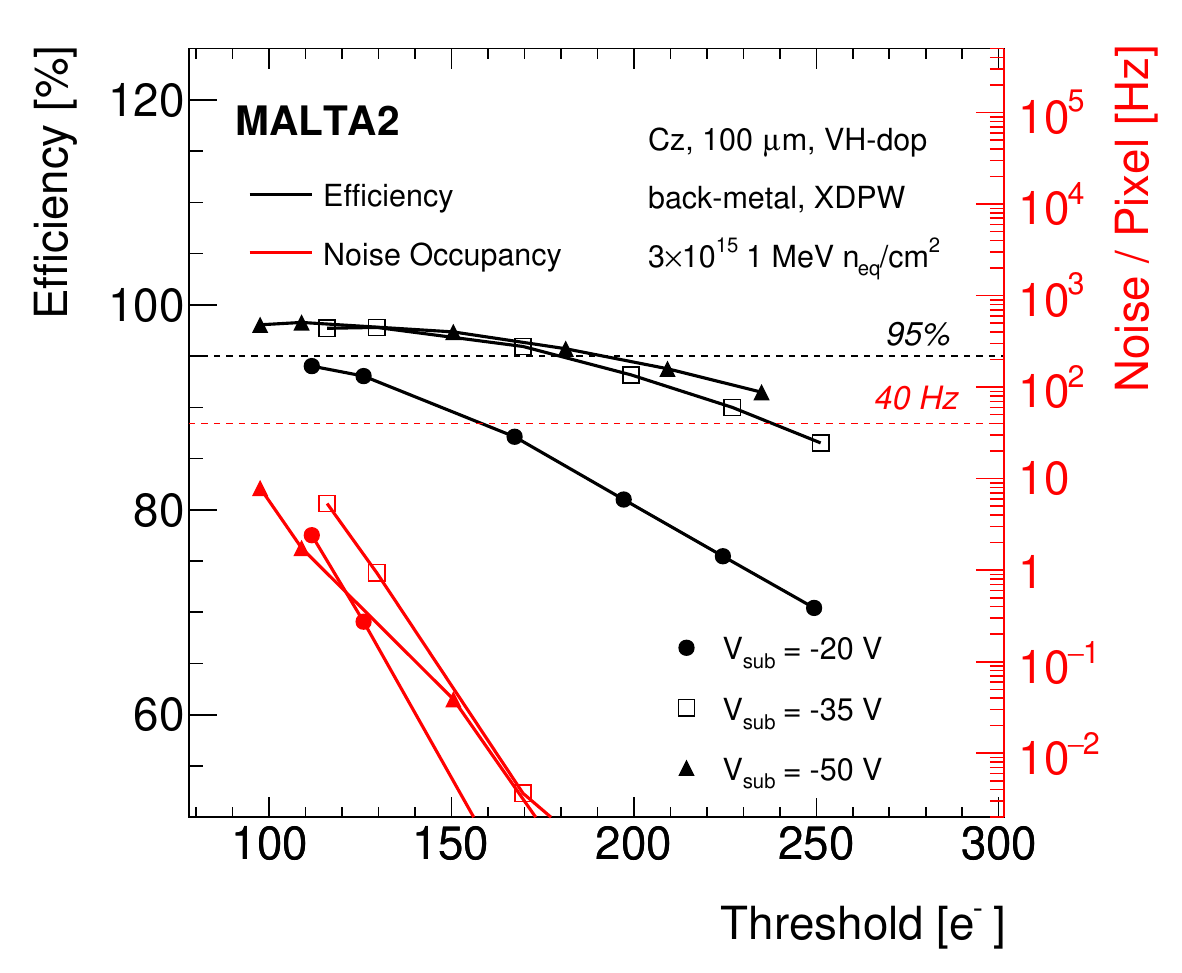}
}
\caption{Average efficiency (in black) and noise occupancy (in red) as a function of threshold in electrons of a MALTA2 sample (Cz, XDPW, very high doping n$^-$ layer, 100 \textmu m, and backside metallisation) irradiated to 3$\times$10$^{15}$ 1 MeV $\mathrm{n_{eq}/{cm}^2}$ . The substrate voltage is set at -20, -35, -50 V. The number of masked pixels is below about 0.02\% of the entire chip in the several configurations.}
\label{fig:vsubs_w18r21}
\end{figure*}

\noindent In order to define an operational window in terms of high efficiency and low noise, the threshold of the in-pixel discriminator and the substrate voltage were systematically varied. The decrease in threshold is expected to increase the number of matched hits, especially for low charge events such as hits in the pixel corners, at the cost of increasing the detected noise. The increase in substrate voltage increases the effective active depth of the sensor, leading to higher efficiency and elevated noise. Additionally, all these parameters are further impacted by the accumulated DDD. As this results in a drop in the measured efficiency and generation-recombination centres that increase the leakage current of the sensor, the sensor noise is enhanced. An operating window for an efficiency $>$95\% (black lines) and noise lower than 40 Hz (red lines) for an irradiated sample at 3$\times$10$^{15}$ 1 MeV $\mathrm{n_{eq}/{cm}^2}$  is found for multiple chip configurations (changing biasing voltage) in Figure \ref{fig:vsubs_w18r21}. The operating window was chosen to comply with the ATLAS ITK requirements \cite{atlas2017technical}.

\section{Timing Performance}
\label{Sec:Timing}

\noindent The timing characterisation of the front-end of the MALTA2 sensor has been reported in Ref.\cite{piro20221}. Due to area constraints of the MALTA2 chip, no Time-over-Threshold (ToT) is directly available from the chip. However, the time-walk of the front-end was measured using analog output monitoring pixels in the matrix. The time-walk of the front-end was measured to be less than 25 ns for 90\% of signals from a $^{90}$Sr source. The charged particles generated by the source create Minimum Ionizing Particles (MIP) like signals with an average charge deposition of approximately 1800 e$^-$. However, signals with a time-walk larger than 25 ns were observed for signals with charge depositions below 200 e$^-$. The time jitter of the front-end electronics was evaluated by charge injection within a pixel using circuitry within the matrix digital readout. The arrival time of hits from the injected charge is compared to the timing of the charge injection trigger pulse transmitted to the chip, by using the PicoTDC with 3 ps binning \cite{perktold2014multichannel}. The time jitter of the MALTA2 front-end electronics was measured to be 0.17 ns for charge injection above 1400 e$^-$, increasing to 4.7 ns at the nominal 100 e$^-$ threshold. \\

\begin{figure}
\centering
\resizebox{0.5\textwidth}{!}{
\includegraphics{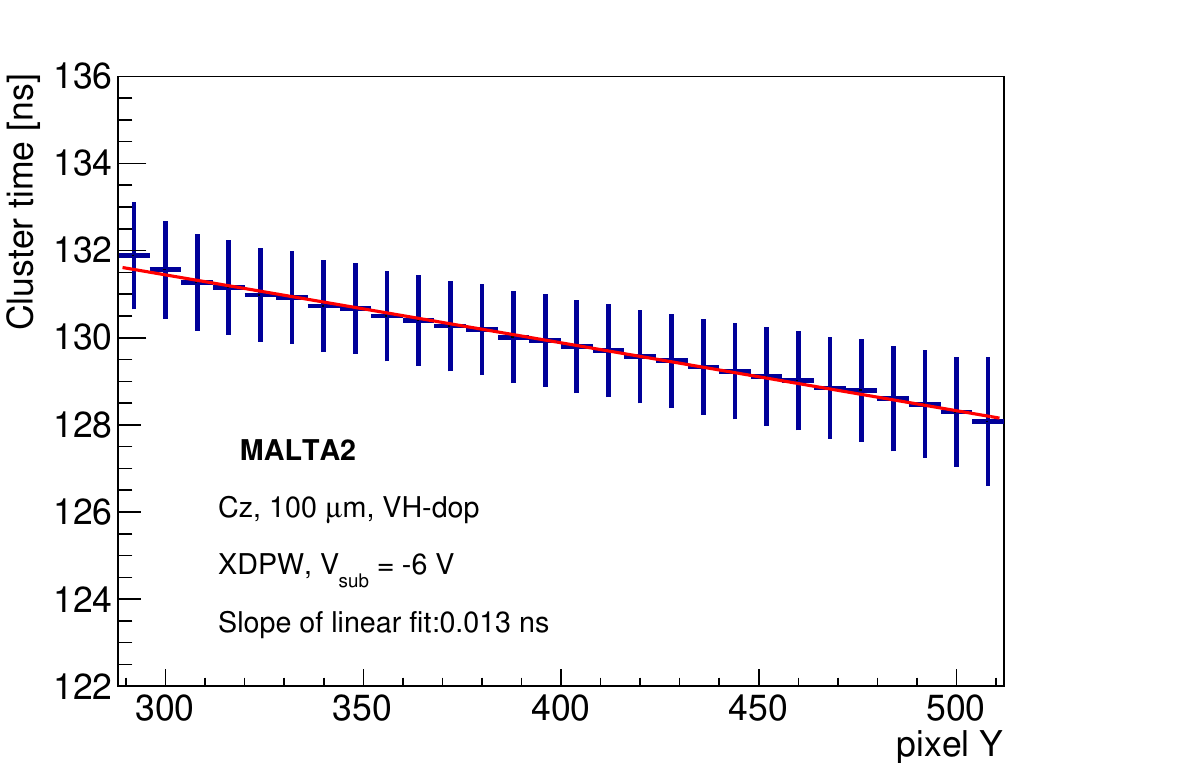}
}
\resizebox{0.5\textwidth}{!}{
\includegraphics{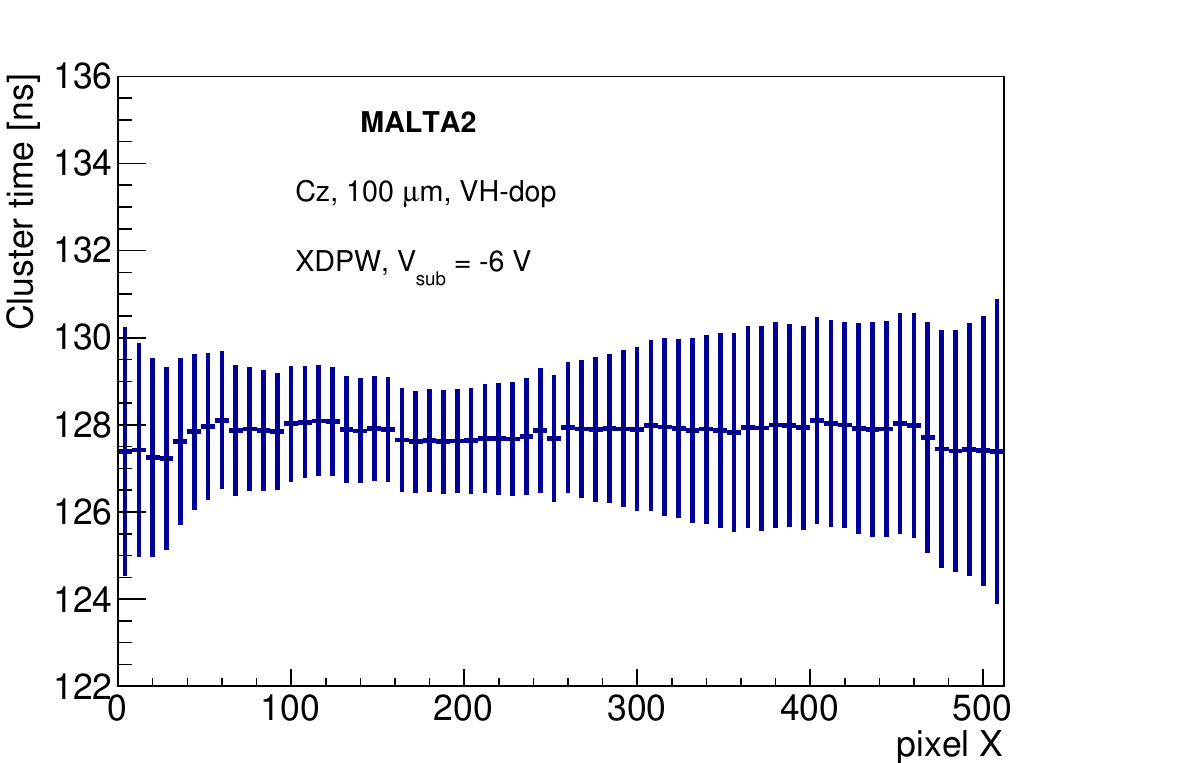}
}
\caption{Time of arrival of leading hit in the cluster with respect to a scintillator reference along the column (left image) and row direction (right image) of the pixel matrix. A correction in the Y-direction is applied due to the time propagation across the column which exhibits a linear behaviour. A correction in the X-direction is applied due to the non-uniformities in the chip response. Error bars represent the corresponding RMS. Both measurements are performed on a MALTA2 sample (Cz, XDPW, very high doping n$^-$ layer, 100 \textmu m thick) at -6 V. Threshold corresponds to 170 e$^-$.}
\label{fig:ClusterTimevsCol}
\end{figure}

\noindent In order to characterise the timing performance of a MALTA2 sample, various intrinsic effects need to be accounted for. As discussed in Ref.\cite{gustavino2022timing}, the time required to reach the periphery along the column direction in the pixel matrix needs to be corrected for. The signal propagation consists of a contribution of the signal generation inside the pixel group and a contribution from the hit propagation to the periphery. As shown in the left image of Figure \ref{fig:ClusterTimevsCol}, this effect exhibits a linear behavior. The error bars represent the corresponding Root Mean Square (RMS). The red line represents the linear fit (with slope of 0.013 ns), which is in turn used as the correcting function on the timing information in the column (Y) direction. Furthermore, a correction is applied across the row (X) direction of the pixel matrix, shown in the right image of Figure \ref{fig:ClusterTimevsCol}. The variation along the X direction is attributed to non-uniformities in the chip response.

\subsection{Before Displacement Damage}

\begin{figure}
\centering
\resizebox{0.4\textwidth}{!}{
\includegraphics{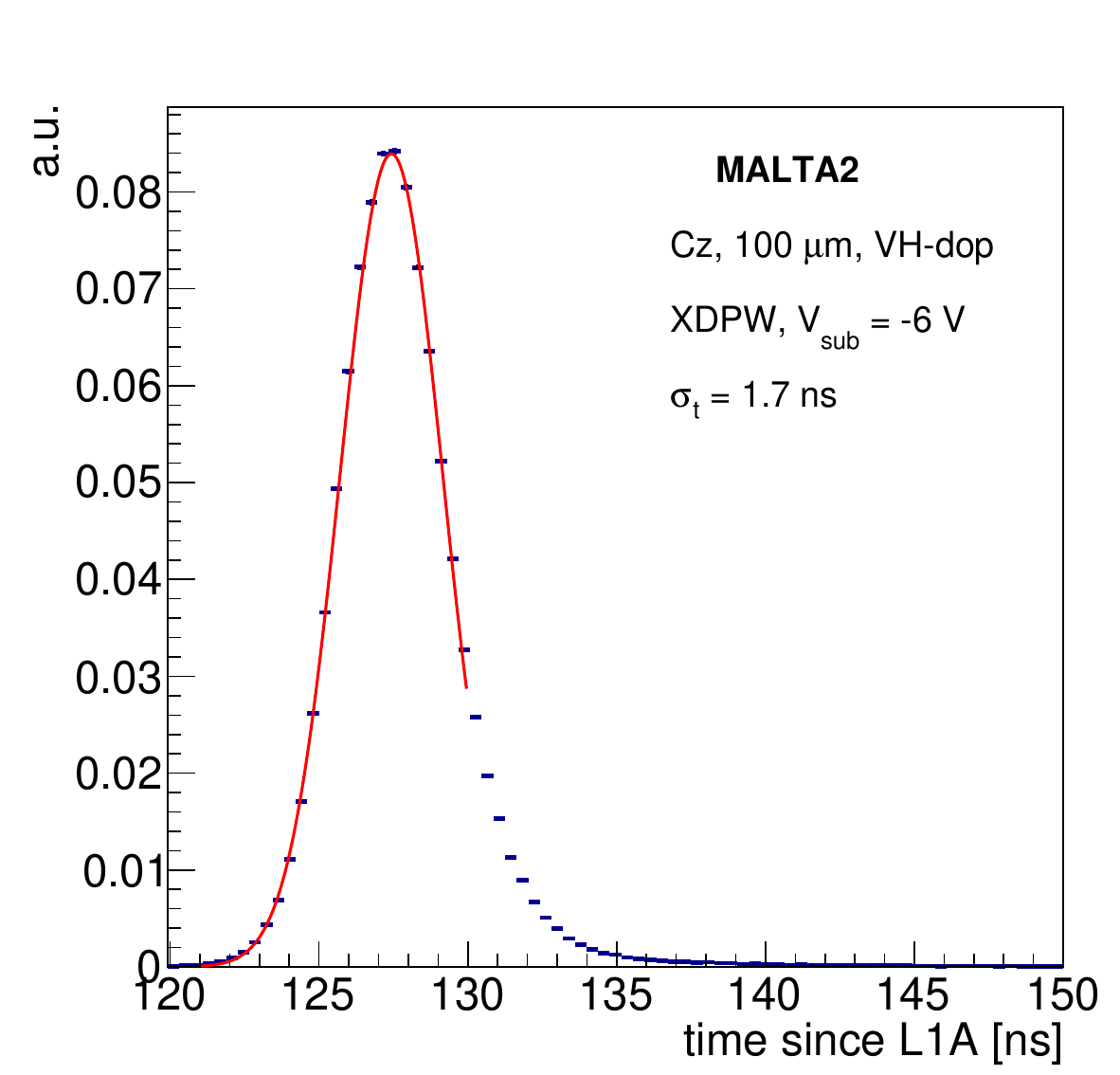}
}
\resizebox{0.4\textwidth}{!}{
\includegraphics{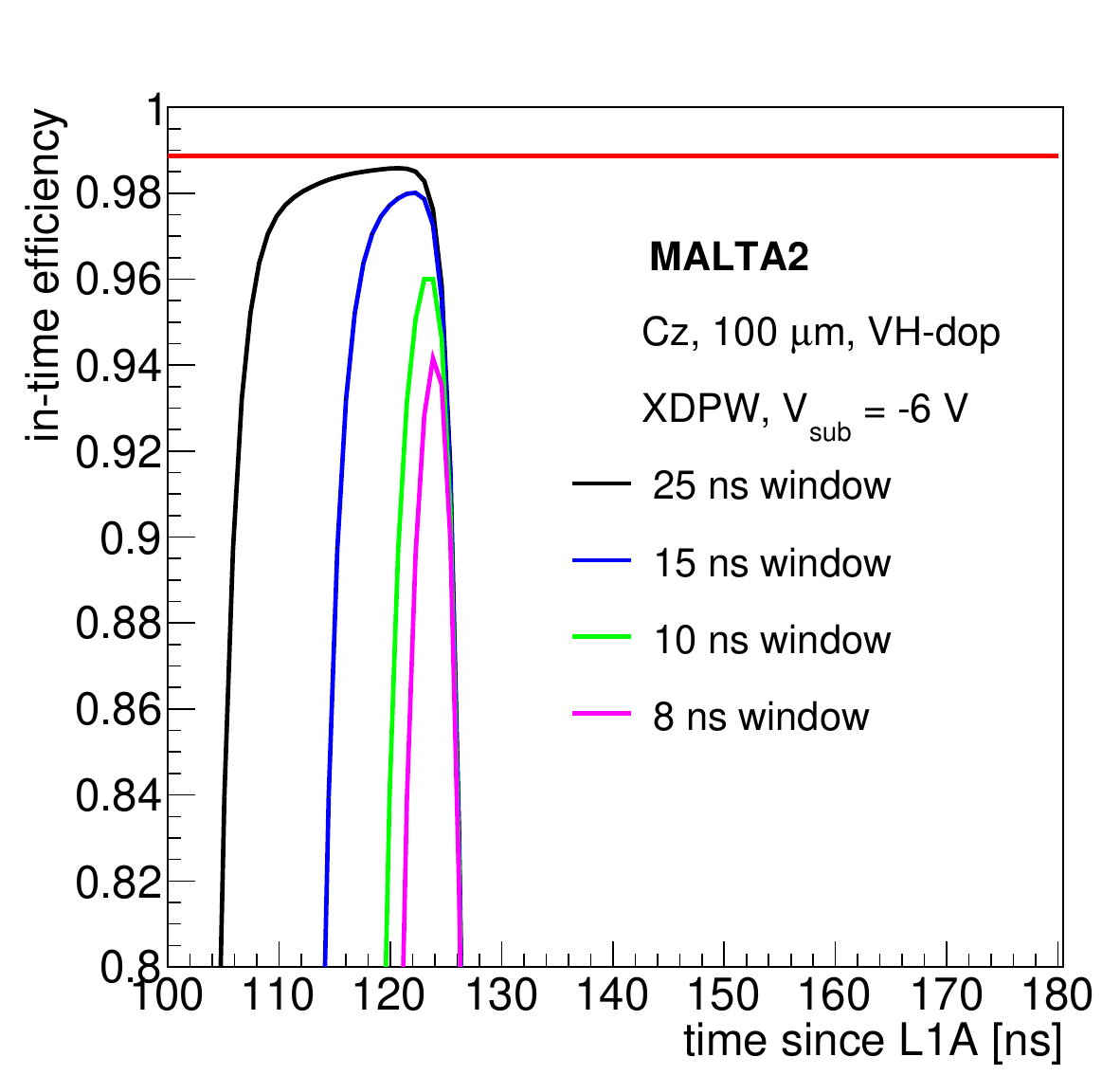}
}
\caption{Left image shows time of arrival of the leading hit in the cluster with respect to a scintillator reference. The quoted $\sigma_t = 1.7$~ns corresponds to the Gaussian fit to the core of the distribution. Right image shows in-time efficiency. Black curve corresponds to in-time efficiency within a 25 ns window, blue curve corresponds to a 15 ns windows, and the green and magenta cure represent a 10 and 8 ns window, respectively. Red line represents maximum achievable efficiency without timing constraints. Both measurements are performed on a MALTA2 sample (Cz, XDPW, very high doping n$^-$ layer, 100 \textmu m thick) at -6 V. Threshold corresponds to 170 e$^-$.}
\label{fig:ToA}
\end{figure}

\noindent The overall MALTA2 timing performance can be assessed by correcting for the aforementioned intrinsic effects and including existing external effects. The timing performance has been measured during the test beam campaigns for both non-irradiated and irradiated chips where a scintillator is used for timing reference. The left image of Figure \ref{fig:ToA} shows the time of arrival of the fastest hit in a pixel cluster with respect to the scintillator reference. The performance is tested on a Cz MALTA2 chip at -6 V bias voltage at a threshold value corresponding to 170~e$^-$. The timing resolution equates to $\sigma$$_{t}=1.7$~ns and is obtained by fitting a Gaussian to the core of the time difference distribution. The distribution contains a jitter contribution from the scintillator ($\mathrm{\sigma_{scintillator}}$$\sim0.5$~ns) and from oversampling within the FPGA ($\mathrm{\sigma_{FPGA}}=3.125/\sqrt{12}=0.9$~ns). For HEP applications such as the LHC, sensor signals need to be registered within the bunch-crossing clock of 25 ns. The in-time efficiency for samples was determined by integrating the time-of-arrival distributions (with respect to the scintillator reference) with a sliding window algorithm. The results in the right image of Figure \ref{fig:ToA} show that for non-irradiated MALTA2 Cz, above 98\% of the hits are collected within a 25 ns time window (black) and 90\% of the hits are collected within 8 ns (magenta), making it suitable for applications at the HL-LHC and other proposed future collider facilities. \\

\subsection{After Displacement Damage}

\begin{figure*}
\centering
\hspace*{3.75cm} 
\resizebox{0.9\textwidth}{!}{
\includegraphics{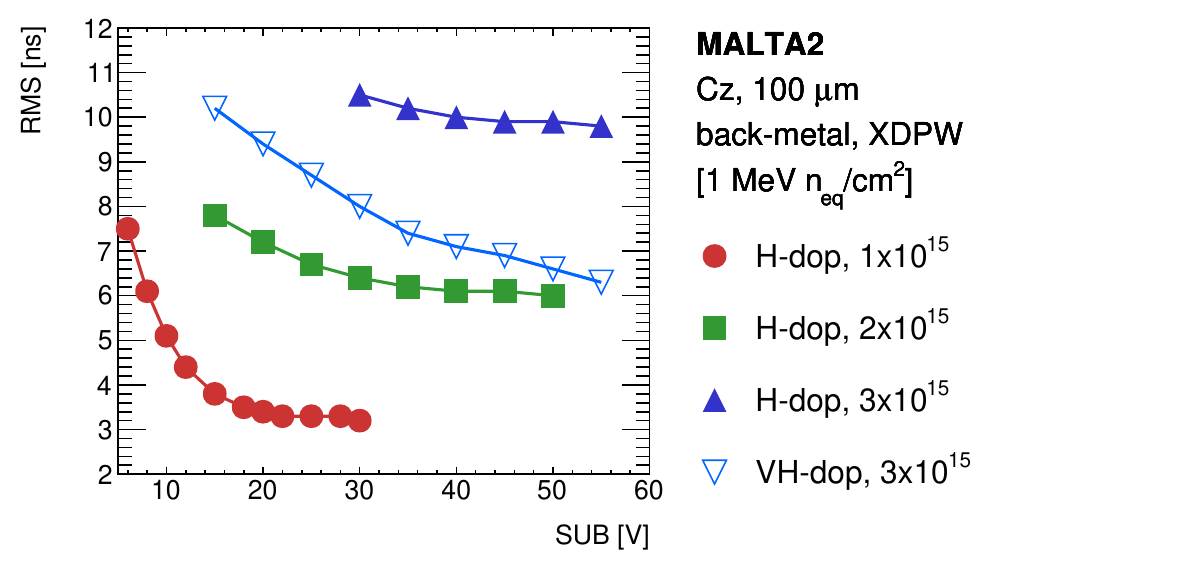}
}
\caption{RMS of timing difference distribution versus bias voltage. Only data points where the detection efficiency lies above 85\% are taken into account. Shown here are four MALTA2 samples (XDPW, 100 \textmu m thick, and backside metallisation). Three samples feature the high doping of the n$^-$ layer. They are irradiated to 1, 2, and 3$\times$10$^{15}$ 1 MeV $\mathrm{n_{eq}/{cm}^2}$  and the results are shown at best performing operating threshold, corresponding to 240, 260, and 120 e$^-$, respectively. One sample features the very high doping of the n$^-$ layer. It is irradiated to 3$\times$10$^{15}$ 1 MeV $\mathrm{n_{eq}/{cm}^2}$  and the results are shown at best performing operating threshold, corresponding to 110 e$^-$.}
\label{fig:RMSvsSUB}
\end{figure*}

 As mentioned in the previous chapter, the bias voltage of irradiated samples can be increased further, which allows to achieve good timing performance for samples irradiated at 1, 2, and 3$\times$10$^{15}$ 1 MeV $\mathrm{n_{eq}/{cm}^2}$ . Figure \ref{fig:RMSvsSUB} shows the evolution of the RMS of the time difference distribution between the leading hit in the cluster and the scintillator reference as a function of bias voltage for irradiated MALTA2 samples with high and very high of the n$^-$ layer. As the timing difference distribution of irradiated samples exhibits a significant tail, a Gaussian fit is not employed, and instead, the RMS is extracted directly from the distribution. Only the cases where the detection efficiency lies higher than 85\% are considered (shown in Figure \ref{fig:EffvsSUB}). The results show that for all samples the RMS decreases as the bias voltage is increased. As the substrate voltage and the depleted area increase, a sharper signal pulse is generated. This faster signal with higher amplitude allows for narrower time-difference distributions. The timing performance is counteracted by the effect of the radiation damage, as the individual timing distributions get broader as the fluence level of the sample increases. This is attributed to various effects, such as charge trapping and changing mobility of charge carriers. The results agree with the trend observed in Figure \ref{fig:ClusSizevsSUB}, where the increase of the bias voltage results in a larger degree of charge sharing. For a sample irradiated to 3$\times$10$^{15}$ 1 MeV $\mathrm{n_{eq}/{cm}^2}$ , the very high doping of the n$^-$ layer significantly helps to improve the RMS of the time difference distribution over a larger range of bias voltages, discussed in more detail in Section \ref{timing_nimplant}. \\

\begin{figure}
\centering
\resizebox{0.5\textwidth}{!}{
\includegraphics{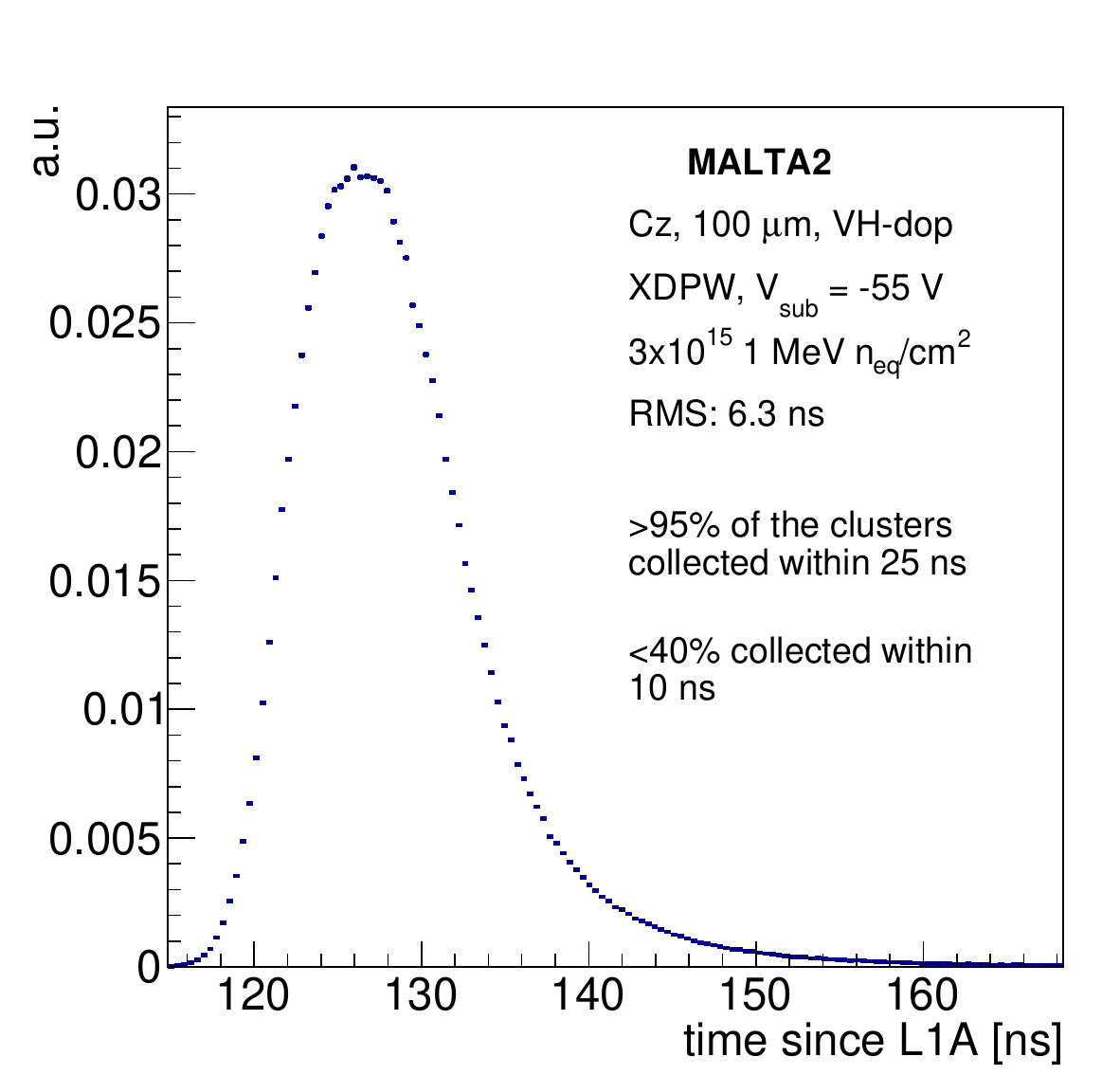}
}
\caption{Time of arrival of leading hit in the cluster with respect to a scintillator reference. Timing measurements are performed on a MALTA2 sample (Cz, XDPW, very high doping of n$^-$ layer, 100 \textmu m thick, and backside metallisation) irradiated to 3$\times$10$^{15}$ 1 MeV $\mathrm{n_{eq}/{cm}^2}$  and operated at -55 V. Threshold corresponds to 110 e$^-$.}
\label{fig:InTimeEff_BackMet}
\end{figure}

\noindent Figure \ref{fig:InTimeEff_BackMet} shows the timing difference distribution of an irradiated (3$\times$10$^{15}$ 1 MeV $\mathrm{n_{eq}/{cm}^2}$ ) MALTA2 Cz with very high doping of the n$^-$ layer. The RMS of the distribution corresponds to 6.3 ns. When using the sliding window in-time efficiency algorithm, as used for the right image of Figure \ref{fig:ToA}, it is found that still more than 95\% of the cluster are collected within 25 ns, whereas less than 40\% of the clusters are collected within a 10 ns window. \\

\begin{figure*}
\centering
\resizebox{1\textwidth}{!}{
\includegraphics{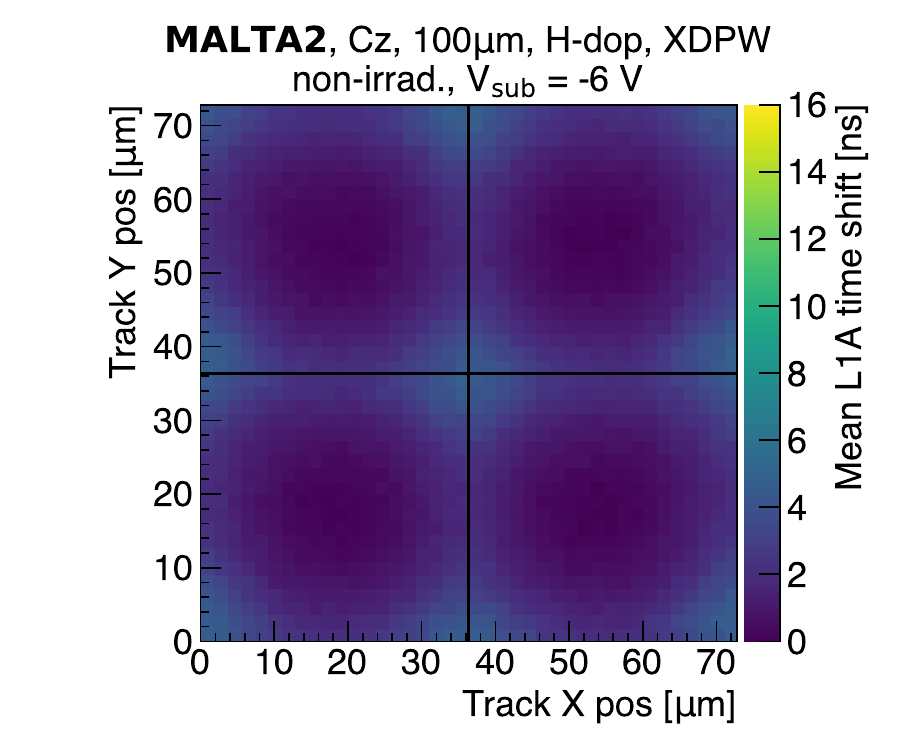}
\includegraphics{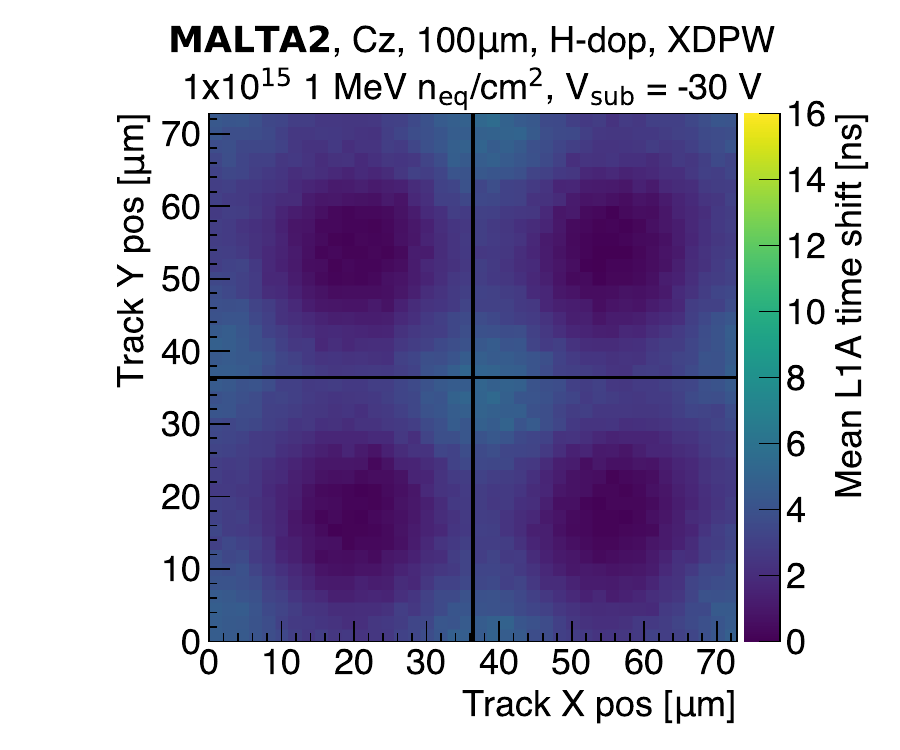}
}
\resizebox{1\textwidth}{!}{
\includegraphics{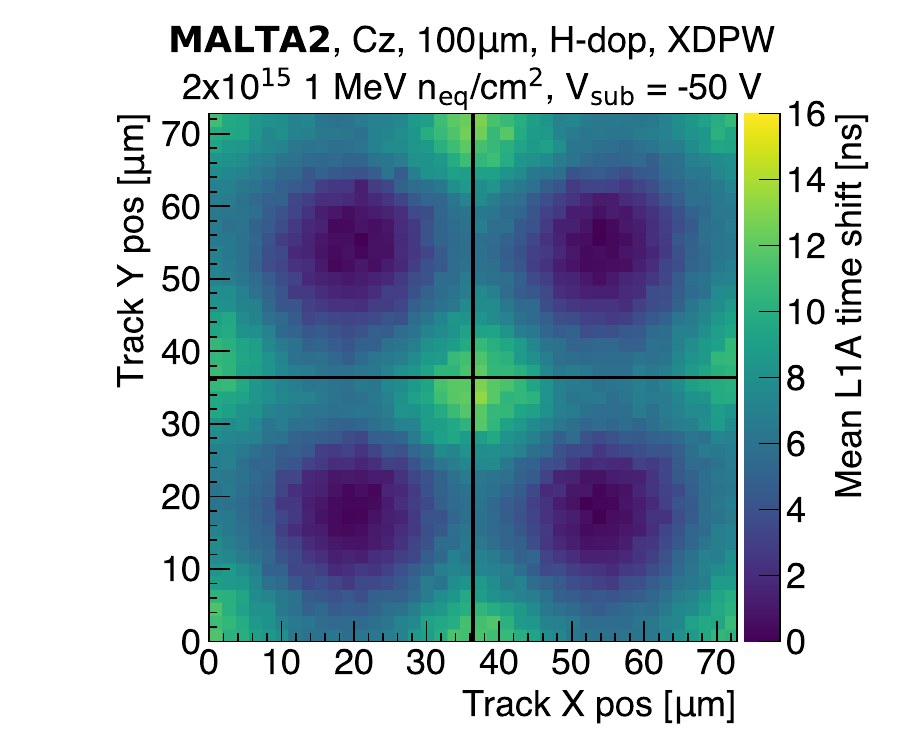}
\includegraphics{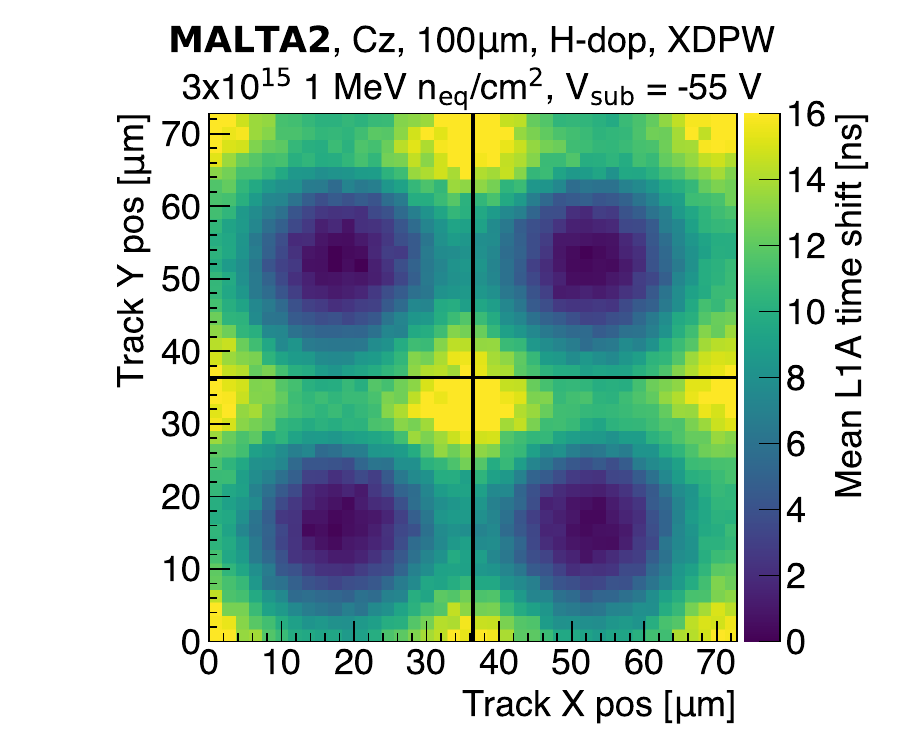}
}
\caption{Projection of the variation of the mean timing of the leading hit within a cluster with respect to a scintillator reference for four MALTA2 samples (Cz, XDPW, high doping of n$^-$ layer, 100 \textmu m thick, and backside metallisation). The samples are irradiated to four different irradiation levels (non-irradiated, 1, 2, and 3$\times$10$^{15}$ 1 MeV $\mathrm{n_{eq}/{cm}^2}$ ) and are operated at best operating threshold, corresponding to 250, 240, 250, and 120 e$^-$, at -6, -30, -50, -55V bias voltage, respectively. The leading hit time data are sorted into 1.82$\times$1.82 \textmu m$^2$ bins based on their associated track position within the pixel extracted from the telescope data. Hits from over the entire chip are projected onto a 2$\times$2 pixel matrix. The quoted mean time value is extracted from a Gaussian fit to the core of the timing distribution for each bin relative to the bin with the smallest value. The operating conditions of the four samples correspond to the data point where the timing RMS is minimised, while the efficiency lies above 90\%.}
\label{fig:2dinpixel}
\end{figure*}

\begin{figure*}
\centering
\hspace*{3.75cm} 
\resizebox{0.9\textwidth}{!}{
\includegraphics{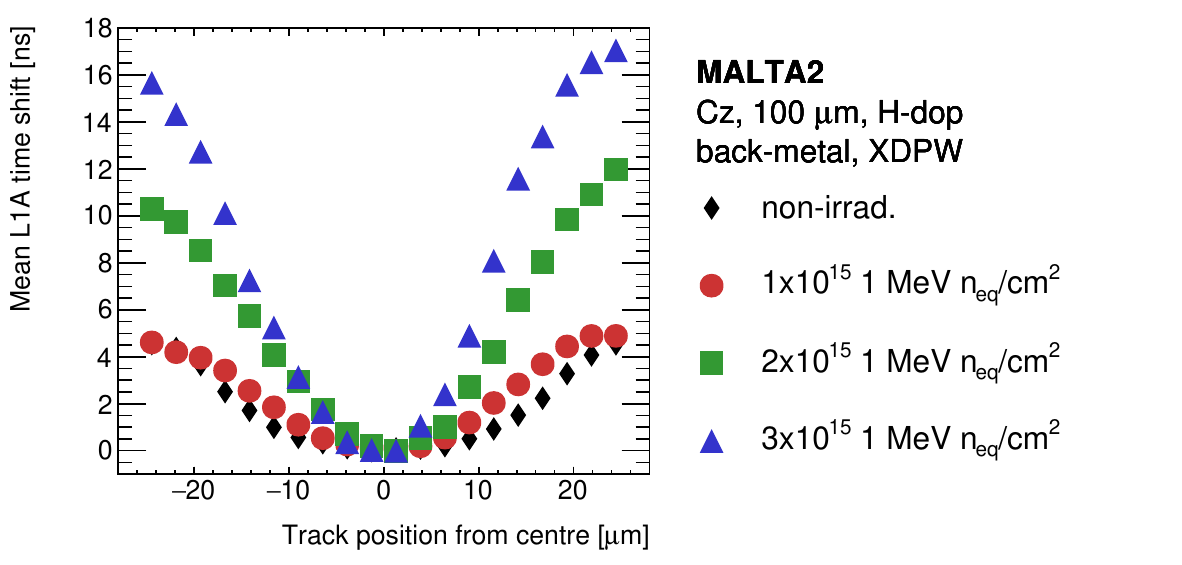}
}
\caption{
Relative shift of the mean time of arrival of the leading hit within cluster with respect to the scintillator reference, as a function of the distance of the associated telescope track from the pixel centre. Results are shown for four MALTA2 samples (Cz, XDPW, high doping of n$^-$ layer, 100 \textmu m thick, and backside metallisation). The samples are irradiated to four different irradiation levels (non-irradiated, 1, 2, and 3$\times$10$^{15}$ 1 MeV $\mathrm{n_{eq}/{cm}^2}$ ) and are operated at best operating threshold, corresponding to 250, 240, 250, and 120 e$^-$, at -6, -30, -50, -55V bias voltage, respectively. The operating threshold point for each sample is selected to minimise the RMS of its timing distribution. The leading hit time data are sorted into 1.82$\times$1.82 \textmu m$^2$ bins based on their associated track position within the pixel. The quoted mean corresponds to the Gaussian fit to the core of the distribution for each bin along the diagonal of the pixel. 
}
\label{fig:diagonalw12}
\end{figure*}

\noindent In order to better understand the impact of irradiation on the samples' timing RMS, the hit tracks are projected onto a 2$\times$2 pixel matrix. The mean timing of the leading hit within a cluster with respect to a scintillator reference varies depending on the region of the pixel being hit. Figure \ref{fig:2dinpixel} shows the difference between the mean time shift of the leading hit in the cluster across the MALTA2 pixel. Comparing the individual figures for four different irradiation levels (non-irradiated, 1, 2, and 3$\times$10$^{15}$ 1 MeV $\mathrm{n_{eq}/{cm}^2}$ ), qualitatively shows that the uniformity of the mean time of arrival of the leading hit deteriorates as the radiation dose increases. Whereas this is not the only effect impacting the RMS value, the loss of uniformity does further increase the timing RMS values as shown in Figure \ref{fig:RMSvsSUB}. To evaluate this effect in a quantitative manner, the bins along the diagonal of the MALTA2 pixel are considered for the samples shown in Figure \ref{fig:2dinpixel}. Considering the bins along the MALTA2 pixel diagonal, Figure \ref{fig:diagonalw12} shows the relative shift of the mean time of hit as a function of the associated telescope track distance from the pixel centre. The quoted mean corresponds to the Gaussian fit to the core of the distribution for each bin along the diagonal of the pixel.\\

\noindent Figure \ref{fig:diagonalw12} demonstrates that the difference between the mean time of hit in the pixel corner and its centre is less than 5 ns in the case of a non-irradiated sample. Only a small change is observed at the dose of 1$\times$10$^{15}$ 1 MeV $\mathrm{n_{eq}/{cm}^2}$  compared to the non-irradiated case. This shift, however, grows with the irradiation dose up to more than three times the original value, reaching around 17 ns in the case of the 3$\times$10$^{15}$ 1 MeV $\mathrm{n_{eq}/{cm}^2}$  sample. Further investigation has shown that the operating threshold point does not have a strong effect on the mean time shift. This observation also suggests that the variation between the samples is not due to the time walk effect. The small ($\sim$2 ns) asymmetry between arbitrarily defined positive and negative distance of the tracks from the pixel centre may stem from the non-exact symmetry of the MALTA2 pixel. Apart from the diagonally asymmetric PWELL, there are additional readout effects that can account for the absence of a perfect mirror symmetry.\\

\subsection{Effect of the Doping Level of the n$^-$ Layer}
\label{timing_nimplant}

\begin{figure}
\centering
\resizebox{0.5\textwidth}{!}{
\includegraphics{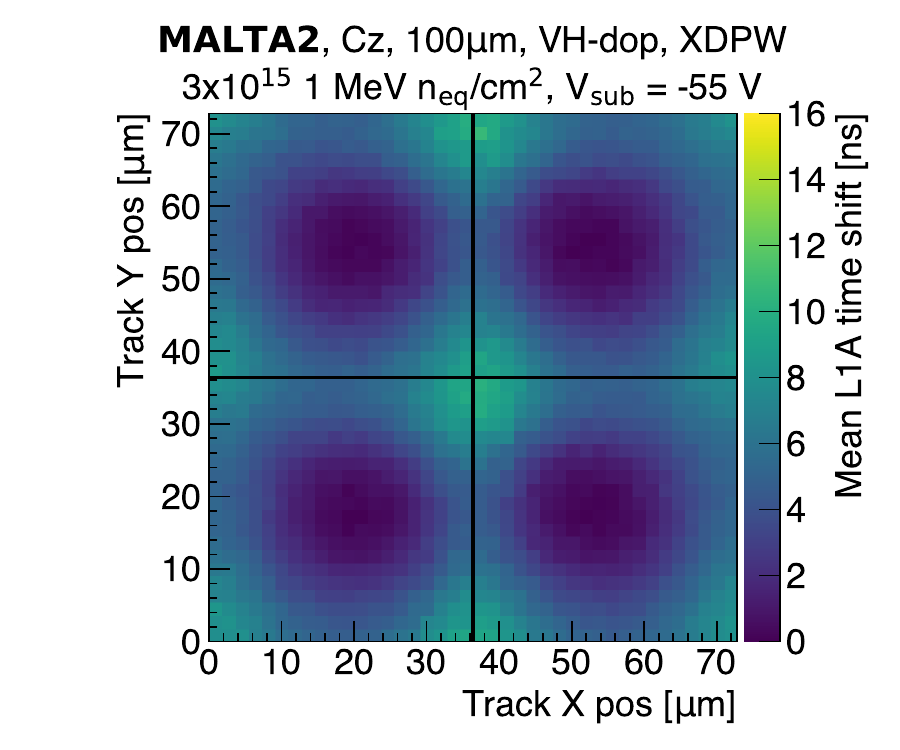}
}
\caption{
Projection of the variation of the mean timing of the leading hit within a cluster with respect to a scintillator reference, for a MALTA2 sample (Cz, XDPW, very high doping of n$^-$ layer, 100 \textmu m thick, and backside metallisation) irradiated to 3$\times$10$^{15}$ 1 MeV $\mathrm{n_{eq}/{cm}^2}$  and operated at $-55$ V. The operating threshold corresponds to $\sim$110 e$^-$.
}
\label{fig:2dinpixelVH}
\end{figure}

\begin{figure*}
\centering
\hspace*{3.75cm} 
\resizebox{0.9\textwidth}{!}{
\includegraphics{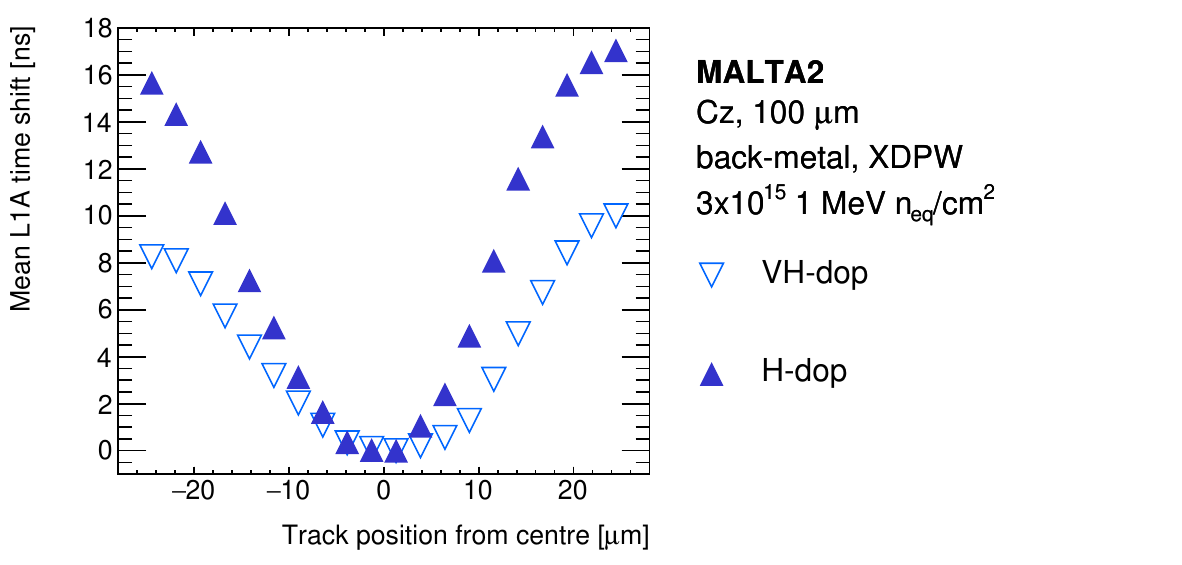}
}
\caption{
Relative shift of the mean time of arrival of the leading hit within a cluster with respect to a scintillator reference, as a function of the distance of the associated telescope track from
the pixel centre. Results are shown for two MALTA2 samples (Cz, XDPW, 100 \textmu m thick, and backside metallisation) irradiated to 3$\times$10$^{15}$ 1 MeV $\mathrm{n_{eq}/{cm}^2}$  and operated at -55V. The samples differ in the doping level of the n$^-$ layer, i.e. high and very high, and the results are shown for best performing operational threshold, corresponding to 120 and 110 e$^-$, respectively. The operating threshold point for both samples is selected to minimise the RMS of its timing distribution.
}
\label{fig:diagonal3e15}
\end{figure*}

\noindent This method is further employed to compare the effect of the doping level of the n$^-$ layer on the time shift at an irradiation level of 3$\times$10$^{15}$ 1 MeV $\mathrm{n_{eq}/{cm}^2}$ . Analogous to Figure \ref{fig:2dinpixel}, the difference between the mean time arrival shift of the leading hit in the cluster with respect to a scintillator reference for a chip with very high doping n$^-$ layer at 3$\times$10$^{15}$ 1 MeV $\mathrm{n_{eq}/{cm}^2}$  is shown in Figure \ref{fig:2dinpixelVH}. The comparison of the mean time shift along the diagonal of the two samples irradiated to the dose 3$\times$10$^{15}$ 1 MeV $\mathrm{n_{eq}/{cm}^2}$  is illustrated in Figure \ref{fig:diagonal3e15}.\\

\noindent Using Figure \ref{fig:diagonal3e15}, it can be shown that out of the two MALTA2 Cz samples irradiated to the fluence of 3$\times$10$^{15}$ 1 MeV $\mathrm{n_{eq}/{cm}^2}$ , the sample with very high doping of the n$^-$ layer demonstrates greater homogeneity of the mean time of hit along the pixel diagonal. As the samples otherwise share identical design parameters, this may indicate that the sample with very high doping of the n$^-$ layer exhibits better radiation tolerance compared to the one with high doping. Despite receiving a dose of 3$\times$10$^{15}$ 1 MeV $\mathrm{n_{eq}/{cm}^2}$ , the very high doping n$^-$ layer sample shows comparable behaviour to a high doping n$^-$ layer sample irradiated to 2$\times$10$^{15}$ 1 MeV $\mathrm{n_{eq}/{cm}^2}$ .

\section{Conclusion}

The MALTA2 monolithic CMOS sensor is the latest DMAPS prototype of the MALTA family. The presented combination of pixel design, process and design modifications, and using high-resistivity Czochralski substrates with backside metallisation for MALTA2 have allowed to explore its performance at NIEL radiation levels up to 3$\times$10$^{15}$ 1 MeV $\mathrm{n_{eq}/{cm}^2}$. Non-irradiated MALTA2 samples on Czochralski substrates can achieve efficiencies of 99\% and an average cluster size of 2 pixels at low threshold settings (150 e$^-$). In these conditions, a timing resolution of $\sigma_t$=1.7 ns can be obtained, where more than 98\% of the hits are collected within 25 ns. Superior performance at the highest irradiation dose (3$\times$10$^{15}$ 1 MeV $\mathrm{n_{eq}/{cm}^2}$ ) was found on samples with very high doping of the n$^-$ layer. Here, a maximum efficiency of 98\% and an average cluster size of 1.7 pixels could be obtained at an operating threshold of 110 e$^-$. At these operating conditions, the RMS of the time difference distribution equals to 6.3 ns, in which 95\% of the clusters are collected within 25 ns. Additionally, the irradiated MALTA2 sample with very high doping of the n$^-$ layer exhibits a more uniform timing response across its pixel compared to a sample with high doping of the n$^-$ layer irradiated to the same fluence. The continuous improvements and adaptations of the MALTA sensor have paved the way for enhanced radiation tolerance and improved performance. As the field of high-energy physics progresses, the experiences and lessons learned from MALTA will undoubtedly contribute to the development of future detectors, pushing the boundaries of scientific frontiers even further.

\begin{acknowledgement}

We are grateful to Grégory Grosset and his colleagues of Ion Beam Services for their support during the backside metallisation process. This project has received funding from the European Union’s Horizon 2020 Research and Innovation programme under Grant Agreement numbers 101004761 (AIDAinnova), 675587 (STREAM), and 654168 (IJS, Ljubljana, Slovenia).

\end{acknowledgement}

\end{document}